\documentclass[sigconf]{acmart}
\usepackage{microtype}
\usepackage{graphicx}
\usepackage{subfigure}
\usepackage{balance}
\usepackage{booktabs} 
\usepackage{amsthm}
\usepackage{array}
\usepackage{graphicx}
\usepackage{clrscode}
\usepackage{subfigure}
\usepackage{multirow}
\usepackage{multicol}
\usepackage{float}
\usepackage{color}
\usepackage{xcolor}
\usepackage{amsopn}
\usepackage{mathrsfs}
\usepackage{mathtools}
\usepackage{amsmath}
\usepackage{booktabs}
\usepackage{arydshln}
\usepackage{hyperref}
\usepackage{blkarray}
\usepackage{enumerate}
\usepackage{courier}
\usepackage{mathrsfs}
\usepackage{rotating}
\usepackage{bm}
\usepackage{wrapfig}
\usepackage{subfigure}
\usepackage{array}
\usepackage{ragged2e}
\usepackage{hyperref}
\usepackage{amsmath}
\usepackage{makecell}
\usepackage{threeparttable}
\usepackage{listings}
\usepackage{xspace}

\usepackage{adjustbox} 
\usepackage{enumitem}

\usepackage{comment}
\usepackage[ruled,linesnumbered]{algorithm2e}

\newcommand{\ie}{\emph{i.e.,}\xspace}
\newcommand{\eg}{\emph{e.g.,}\xspace}

\newcommand{\ignore}[1]{}

\SetKwComment{comment}{ $triangleright$ \ }{}

\theoremstyle{definition}

\theoremstyle{theorem}

\theoremstyle{proof}

\theoremstyle{remark}

\AtBeginDocument{%
  \providecommand\BibTeX{{%
    \normalfont B\kern-0.5em{\scshape i\kern-0.25em b}\kern-0.8em\TeX}}}

\setcopyright{acmcopyright}
\copyrightyear{2021}
\acmYear{2021}
\acmDOI{10.1145/1122445.1122456}

\acmConference[Woodstock '21]{Woodstock '18: ACM Symposium on Neural
  Gaze Detection}{June 03--05, 2018}{Woodstock, NY}
\acmBooktitle{Woodstock '21: ACM Symposium on Neural Gaze Detection,
  June 03--05, 2021, Woodstock, NY}
\acmPrice{15.00}
\acmISBN{978-1-4503-XXXX-X/18/06}

\begin{document}
\title[RecBole 2.0: Towards a More Up-to-Date Recommendation Library]{\texorpdfstring{RecBole 2.0: Towards a More Up-to-Date \\ Recommendation Library}{Resource: RecBole 2.0: Towards a More Up-to-Date  Recommendation Library}}

\author{Wayne Xin Zhao$^{1,2}$, Yupeng Hou$^{1,2,\#}$, Xingyu Pan$^{1,3,\#}$, Chen Yang$^{1,2}$}
\author{Zeyu Zhang$^{1}$, Zihan Lin$^{1,3}$, Jingsen Zhang$^{1,3}$, Shuqing Bian$^{1,3}$, Jiakai Tang$^{1}$}
\author{Wenqi Sun$^{1,2}$, Yushuo Chen$^{1,2}$, Lanling Xu$^{1}$, Gaowei Zhang$^{1}$, Zhen Tian$^{1}$}
\author{Changxin Tian$^{1,3}$, Shanlei Mu$^{1,3}$, Xinyan Fan$^{1,2}$, Xu Chen$^{1,2,*}$ and Ji-Rong Wen$^{1,2,3}$}
\thanks{$*$ Xu Chen (successcx@gmail.com) is the corresponding author.}
\thanks{$\#$ Both authors contributed equally to this work.}
\affiliation{\institution{$^1$Beijing Key Laboratory of Big Data Management and Analysis Methods} \country{}} 
\affiliation{\institution{\{$^2$Gaoling School of Artificial Intelligence, $^3$School of Information\} Renmin University of China} \country{}}

\renewcommand{\authors}{Wayne Xin Zhao, Yupeng Hou, Xingyu Pan, Chen Yang, Zeyu Zhang, Zihan Lin, Jingsen Zhang, Shuqing Bian, Jiakai Tang, Wenqi Sun, Yushuo Chen, Lanling Xu, Gaowei Zhang, Zhen Tian, Changxin Tian, Shanlei Mu, Xinyan Fan, Xu Chen and Ji-Rong Wen}
\renewcommand{\shortauthors}{Zhao, et al.}

\begin{abstract}
In order to support the study of recent advances in recommender systems, this paper presents an extended recommendation library consisting of eight packages for up-to-date topics and architectures. First of all, from a data perspective, we consider three important topics related to data issues (\ie  \emph{sparsity}, \emph{bias} and \emph{distribution shift}), and develop five  packages accordingly:
meta-learning, data augmentation, debiasing, fairness and cross-domain recommendation. Furthermore, from a model perspective, we develop two benchmarking packages for Transformer-based and graph neural network~(GNN)-based models, respectively. All the  packages (consisting of 65 new models) are developed based on a popular recommendation framework RecBole, ensuring that both the implementation and interface are unified. 
For each package, we provide complete implementations from data loading, experimental setup, evaluation and algorithm implementation. 
This library provides a valuable resource to facilitate the up-to-date research in recommender systems. The project is released at the link: \url{https://github.com/RUCAIBox/RecBole2.0}.
\end{abstract}

\keywords{Recommendation library;
Reproducibility;
Evaluation}
\maketitle

\section{Introduction}
Nowadays, recommender systems have deeply revolutionized people's daily life, bringing a huge amount of business value and great convenience for information seeking.
In the literature, various recommendation algorithms have been proposed based on different architectures or approaches~\cite{rs_survey}. 
Despite the great progress in recommender systems, there are increasing concerns on the reproducibility of the recommendation algorithms~\cite{Elliot,zhao2021recbole}. Facing with this issue, a number of open-sourced recommendation libraries have been released to facilitate the  reproducible implementation of the proposed recommendation algorithms~\cite{zhao2021recbole,Elliot,Chorus,DaisyRec,bars,microsoft_rec,case_rec,beta-rec,surprise,neurec}. These libraries largely enhance the reproducibility of recommendation algorithms for the research purpose. 

However, existing recommendation libraries mainly focus on classical models,  lacking consideration of the recent advances in the recommendation field, including both new models (\eg graph neural networks~\cite{wu2022gnn4rec_survey}) and new topics (\eg debiasing~\cite{bias_survey} and fairness~\cite{fairness_survey}). 
Due to the rapid progress in recommender systems, we argue that a more up-to-date recommendation library is needed for supporting the research on new advances.
It is particularly important to standardize these ongoing studies at an early stage, preventing non-standard implementation or unreliable evaluation.  


Inspired by this motivation, this paper presents a significant extension of a previously released recommendation library RecBole~\cite{zhao2021recbole}\footnote{https://recbole.io/} (receiving extensive attention of 1.9K stars on  GitHub), by incorporating a series of benchmarking packages for up-to-date advances in recommender systems. In particular, our extension is conducted in two major aspects, namely \emph{data} and \emph{model}. First of all, there is an increasing attention to the issues from the interaction data itself~\cite{chen2020bias}, and we focus on three important research topics related to data issues, namely \emph{data sparsity}, \emph{data bias} and \emph{data distribution shift}. Considering the three data issues, we develop five benchmarking packages corresponding to meta-learning, data augmentation, debiasing, fairness and cross-domain recommendation.   
Furthermore, from a model perspective, we consider providing more support for recommendation algorithms based on emerging model architectures, and develop two benchmarking packages for Transformer based and graph neural network~(GNN) based models.
It is worth noting that these packages (except the Transformer package) provide complete implementations from data loading, experimental setup, evaluation and algorithm implementation, and researchers can quickly produce benchmarking results of a package according to their requirements with little effort.


The merits of this extended library are threefold. First, it is \emph{unified} in implementation and interface:  
 all the packages are developed fully based on the recommendation framework RecBole, reusing existing functions or modules from the main code repository as much as possible.  
Second, it is a \emph{significant} extension with 65 newly implemented algorithms or models, having a good coverage of recent advances in recommender system. With this extension,   RecBole has become the most comprehensive recommendation library in terms of both algorithms (130+ models) and tasks (11 tasks or topics), among the public research projects in recommender systems on GitHub.  
Third, it is \emph{flexible} and \emph{reliable} to use:  we implement the complete evaluation pipeline for each package and carefully conduct the code reviewing and testing. 

\begin{figure}[t]
\centering
\setlength{\fboxrule}{0.pt}
\setlength{\fboxsep}{0.pt}
\fbox{
\includegraphics[width=.9\linewidth]{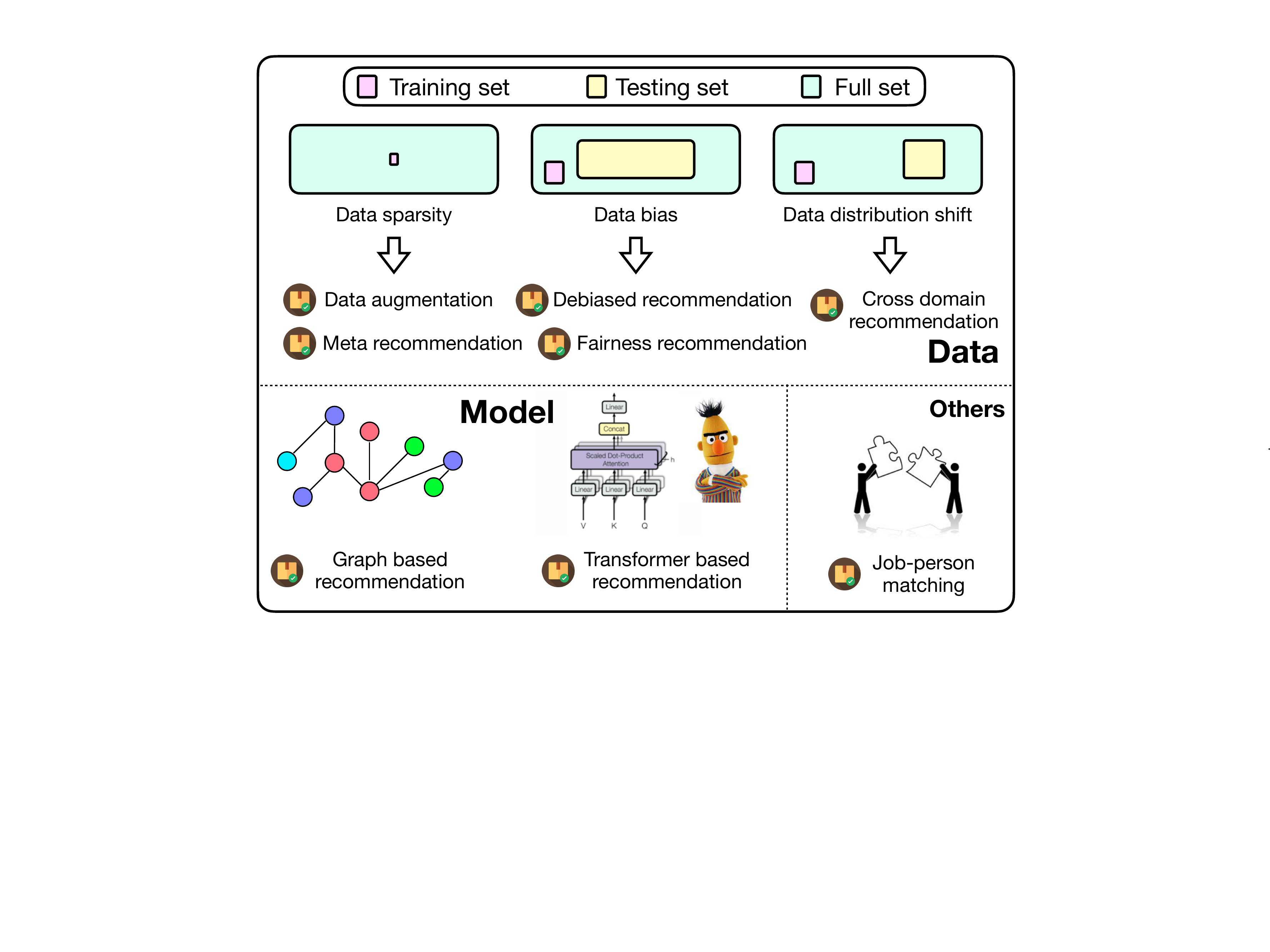}
}
\vspace{-0.2cm}
\caption{Overview of the implemented packages, which are grouped into the data and model parts.}
\vspace{-0.3cm}
\label{intro}
\end{figure}

In the following sections, we first overview the implemented packages, then detail each one to illustrate our design principles and show the usage examples, and finally discuss the extensions of our library as compared with RecBole.

\section{Overview}
We present the overall introduction of the extended library in Figure~\ref{intro}.
As we can see, our extensions consist of eight packages, grouped into   two major parts concerning \emph{data} and \emph{model}. 
For the data packages, we focus on three key issues, that is, data sparsity, data bias and data distribution shift.
To solve the data sparsity issue, we develop two packages, namely data augmentation (generating additional data samples for optimizing recommendation models) and meta-learning based recommendation (approaching the cold-start issue with meta-learning methods).  
To alleviate the data bias, we develop two packages, namely debiased ({reducing the data bias}) and fairness recommendation ({enforcing the fairness of the recommendations}).
To overcome the data distribution shift problem, we implement a cross-domain recommendation package. Note that cross-domain recommendation is not a new research topic, while our focus is to include the most up-to-date cross-domain recommendation models. 
For the model packages, we would like to collect and implement the most up-to-date recommender models.
Specially, we focus on graph neural network and Transformer based models.

Besides, we also implement an application package for the task of person-job fit~\cite{edwards1991person}, which has received much attention from the research community. 
In our library, the optimal configurations (obtained by grid search in predefined parameter ranges) for all the models have been provided, based on which one can easily produce the benchmarking results for each package.
In the following, we introduce the above packages in detail.

\begin{table*}[t]
  \caption{The included packages and the implemented models in each package. }
  \vspace{-0.1cm}
  \center
  \renewcommand\arraystretch{1.2}
  \setlength{\tabcolsep}{5.1pt}
  \begin{threeparttable}  
    \scalebox{.85}{
      \begin{tabular}
        { lm{5cm}<{\centering}m{13cm}<{\centering} m{6cm}<{\centering}
        } \toprule
        Module&Package&Models\\ \hline
        \multirow{8}{*}{Data}&Data augmentation (RecBole-DA)&CL4SRec~\cite{DBLP:journals/corr/abs-2010-14395}, DuoRec~\cite{abs-2110-05730}, MMInfoRe~\cite{QiuHY21}, CauseRec~\cite{ZhangYZC021}, CASR~\cite{WangZXCZZW21}, CCL~\cite{BianZZCHYW21}, CoSeRec~\cite{abs-2108-06479}\\\cline{2-4}
                             &Meta recommendation (RecBole-MetaRec)&MeLU~\cite{lee2019melu}, MAMO~\cite{dong2020mamo}, LWA~\cite{vartak2017meta}, NLBA~\cite{vartak2017meta}, TaNP~\cite{lin2021task}, MetaEmb~\cite{pan2019warm}, MWUF~\cite{zhu2021learning}\\\cline{2-4}
                             &Debiased recommendation (RecBole-Debias)&MF-IPS~\cite{schnabel2016recommendations}, PDA~\cite{zhang2021causal}, MACR~\cite{wei2021model}, DICE~\cite{zheng2021disentangling}, CausE~\cite{bonner2018causal}, Rel-MF~\cite{saito2020unbiased}\\\cline{2-4}
                             &Fairness recommendation (RecBole-FairRec)&{FOCF}~\cite{FOCF}, {PFCN}~\cite{PFCN}, {FairGo}~\cite{FairGo}, {NFCF}~\cite{NFCF}.\\\cline{2-4}
                             &Cross-domain recommendation (RecBole-CDR)&CMF~\cite{cmf}, CLFM~\cite{clfm}, DTCDR~\cite{dtcdr}, DeepAPF~\cite{deepapf}, NATR~\cite{natr}, CoNet~\cite{conet}, BiTGCF~\cite{bitgcf}, EMCDR~\cite{emcdr}, SSCDR~\cite{sscdr} and DCDCSR~\cite{dcdcsr}\\\hline
        \multirow{2}{*}{Model}&{Graph based recommendation (RecBole-GNN)}&NGCF~\cite{wang2019ngcf}, LightGCN~\cite{he2020lightgcn}, SGL~\cite{wu2021sgl}, HMLET~\cite{kong2022hmlet}, NCL~\cite{lin2022ncl}, SimGCL~\cite{yu2022simgcl}, SR-GNN~\cite{wu2019srgnn}, GC-SAN~\cite{xu2019gcsan}, NISER~\cite{gupta2019niser}, LESSR~\cite{chen2020lessr}, TAGNN~\cite{yu2020tagnn}, GCE-GNN~\cite{wang2020gcegnn}, SGNN-HN~\cite{pan2020sgnnhn}, DiffNet~\cite{wu2019diffnet}, MHCN~\cite{yu2021mhcn}, SEPT~\cite{yu2021sept}\\\cline{2-4}
                              &Transformer based recommendation (RecBole-TRM)&
                              TiSASRec~\cite{TiSASRec},
                              SSE-PT~\cite{SSE-PT},
                              LightSANs~\cite{LightSANs},
                              gMLP~\cite{gMLP},
                              CORE~\cite{hou2022core},
                              NRMS~\cite{NRMS}, NAML~\cite{NAML}, NPA~\cite{NPA}
                              \\\hline
 \multirow{1}{*}{Others}&{Person-job fit (RecBole-PJF)}& PJFNN~\cite{zhu2018person}, APJFNN~\cite{qin2018enhancing}, BPJFNN~\cite{qin2018enhancing}, IPJF~\cite{le2019towards}, PJFFF~\cite{jiang2020learning}, SHPJF~\cite{hou2022leveraging}, LFRR~\cite{neve2019latent}\\
                              \bottomrule
      \end{tabular}
    }        
  \end{threeparttable}  
  \vspace{-0.2cm}
  \label{model}   
\end{table*}

\vspace{-0.1cm}
\section{Package Details and Usage}
We organize the packages in two groups, namely data-oriented and model-oriented packages, which are detailed below.

\vspace{-0.1cm}
\subsection{Data-oriented Packages}
$\quad$\textbf{Data augmentation (RecBole-DA)}.
To alleviate the data sparsity problem, a recently proposed technique is to generate additional samples to densify the original user-item interactions~\cite{DBLP:journals/corr/abs-2010-14395,WangZXCZZW21}.
Following \cite{abs-2108-06479,BianZZCHYW21,WangZXCZZW21}, we implement three types of models based on different data augmentation strategies, including 
heuristic methods (CL4SRec~\cite{DBLP:journals/corr/abs-2010-14395} and DuoRec~\cite{abs-2110-05730}), model based methods (MMInfoRec~\cite{QiuHY21} and CauseRec~\cite{ZhangYZC021}) and hybrid methods (CASR~\cite{WangZXCZZW21}, CCL~\cite{BianZZCHYW21} and CoSeRec~\cite{abs-2108-06479}).
Besides providing model implementations, we also provide easy-to-use APIs to configure and combine different data augmentation strategies.

\ignore{
In our library, we implement three types of models following the above data augmentation idea, which includes: the heuristic methods, model based methods and hybrid methods.
More specifically, for heuristic methods, we realize CL4SRec~\cite{DBLP:journals/corr/abs-2010-14395} and DuoRec~\cite{abs-2110-05730}.
For model-based methods, we implement MMInfoRec~\cite{QiuHY21} and CauseRec~\cite{ZhangYZC021}.
For hybird methods, we focus on CASR~\cite{WangZXCZZW21}, CCL~\cite{BianZZCHYW21} and CoSeRec~\cite{abs-2108-06479}. 
It should be noted that we implement the above models through a series of easy-to-use API's, and one can flexibly assemble them to develop new data augmentation methods.
}

\textbf{Meta recommendation (RecBole-MetaRec).} Originated from computer vision and machine learning, meta-learning~\cite{finn2017model,nichol2018first,hospedales2020meta,huisman2021survey} is a principled approach to dealing with few-shot learning tasks. We implement three types of meta-learning recommendation models, for prediction (MeLU~\cite{lee2019melu} and MAMO~\cite{dong2020mamo}), parameterization (LWA~\cite{vartak2017meta}, NLBA~\cite{vartak2017meta} and TaNP~\cite{lin2021task}) and embedding (MetaEmb~\cite{pan2019warm} and MWUF~\cite{zhu2021learning}). These models are implemented by a series of general modules (MetaDataset, MetaDataLoader, MetaRecommender, MetaTrainer, MetaCollector and MetaUtils), which are flexible to be extended with new models. 
\ignore{
Meta learning is firstly proposed to solve the few-shot learning problems in computer vision (CV)~\cite{finn2017model,nichol2018first,hospedales2020meta,huisman2021survey}. 
Recently, many researchers migrate the power of meta learning to alleviate the cold start problem in the recommendation domain.
In this package, we implement three types of models:
(1) meta learn to predict, which includes MeLU~\cite{lee2019melu} and MAMO~\cite{dong2020mamo}. 
(2) Meta learn to parameterize, which includes LWA~\cite{vartak2017meta}, NLBA~\cite{vartak2017meta} and TaNP~\cite{lin2021task}. 
(3) Meta learn to embed, which include MetaEmb~\cite{pan2019warm} and MWUF~\cite{zhu2021learning}. 
For better implementing the above models, we extend RecBole with six additional modules including the MetaDataset, MetaDataLoader, MetaRecommender, MetaTrainer, MetaCollector and MetaUtils, which can be seen on our project page.
}

\textbf{Fairness recommendation (RecBole-FairRec).} The second  package related to data bias is targeted at fairness recommendation~\cite{zehlike2021fairness}, considering the data bias from user perspective (debiased recommendation package mainly focuses on the bias from item side). Specifically, we implement four models in this package including {FOCF}~\cite{FOCF}, {PFCN}~\cite{PFCN}, {FairGo}~\cite{FairGo} and {NFCF}~\cite{NFCF}.
Besides models,  we implement a series of fairness metrics, which are particularly important for fairness recommendation, including Gini Index~\cite{FCPO}, Popularity Rate~\cite{FCPO}, Differential Fairness~\cite{NFCF}, Value Unfairness, Absolute Unfairness, Underestimation Unfairness, Overestimation Unfairness~\cite{FOCF} and Non-Parity Unfairness~\cite{NonParity}.

\textbf{Debiased recommendation (RecBole-Debias).}
Data bias ubiquitously exists in the observed user-item interaction data in recommender systems~\cite{chen2020bias}.  
To correct these biases, we implement six debiased models considering selection bias (MF-IPS~\cite{schnabel2016recommendations}), popularity bias (PDA~\cite{zhang2021causal}, MACR~\cite{wei2021model}, DICE~\cite{zheng2021disentangling} and CausE~\cite{bonner2018causal}) and exposure bias (Rel-MF~\cite{saito2020unbiased}). Besides, we also implement specific dataloaders for three major debiasing datasets (Yahoo!R3~\cite{schnabel2016recommendations}, ML-100k~\cite{schnabel2016recommendations} and KuaiRec~\cite{gao2022kuairec}) for conveniently reproducing the experiments. 

\textbf{Cross-domain recommendation  (RecBole-CDR).}
Data distribution shift often occurs in cross-domain recommendation. To conduct effective cross-domain models, we consider three representative categories of methods, including  collective matrix factorization (CMF~\cite{cmf} and CLFM~\cite{clfm}), representation sharing or combination (DTCDR~\cite{dtcdr}, DeepAPF~\cite{deepapf} and NATR~\cite{natr}) and knowledge transfer or mapping (CoNet~\cite{conet}, BiTGCF~\cite{bitgcf}, EMCDR~\cite{emcdr}, SSCDR~\cite{sscdr} and DCDCSR~\cite{dcdcsr}). 
Cross-domain recommendation is actually not a new topic, but few packages have a good coverage of various representative methods in this research line. 

\ignore{
In real-world applications, the user behavior patterns may vary a lot across different domains, which makes the model learned based on one domain hard to be applied to the other domains.
The fundamental problem lies behind this phenomenon is the data distribution shift.
For alleviating this problem, in this package, we implement many cross-domain recommender models.
In specific, we focus on the following three categories:
(1) algorithms based on the collective matrix factorization, such as CMF~\cite{cmf} and CLFM~\cite{clfm}.
(2) Algorithms that share or combine the representations of the overlapped data, for example, DTCDR~\cite{dtcdr}, DeepAPF~\cite{deepapf} and NATR~\cite{natr}.
(3) Algorithms that transfer or map knowledge between different domains, such as CoNet~\cite{conet}, BiTGCF~\cite{bitgcf}, EMCDR~\cite{emcdr}, SSCDR~\cite{sscdr} and DCDCSR~\cite{dcdcsr}.
In addition, we implement the above models following three important rules: (1) compatible and automatic data processing, (2) flexible and customized model training and (3) unified and extensible evaluation.
}

\vspace{-0.cm}
\subsection{Model-oriented Packages}
$\quad$\textbf{GNN based recommendation (RecBole-GNN).}
In recent years, graph neural networks (GNNs)~\cite{kipf2017gcn,hamilton2017graphsage,velickovic2018gat} have shown to be effective to model graph structures of various data types, \eg recommender systems~\cite{wu2022gnn4rec_survey}. 
We implement three types of GNN models tailored to different tasks, including general recommendation (NGCF~\cite{wang2019ngcf}, LightGCN~\cite{he2020lightgcn}, SGL~\cite{wu2021sgl}, HMLET~\cite{kong2022hmlet}, NCL~\cite{lin2022ncl} and SimGCL~\cite{yu2022simgcl}), sequential recommendation (SR-GNN~\cite{wu2019srgnn}, GC-SAN~\cite{xu2019gcsan}, NISER~\cite{gupta2019niser}, LESSR~\cite{chen2020lessr}, TAGNN~\cite{yu2020tagnn}, GCE-GNN~\cite{wang2020gcegnn} and SGNN-HN~\cite{pan2020sgnnhn}) and social recommendation (DiffNet~\cite{wu2019diffnet}, MHCN~\cite{yu2021mhcn} and SEPT~\cite{yu2021sept}). Following the original data mechanism, we design a new kind of atomic file with suffix \texttt{.net} for structured data (\emph{e.g.}, social network) modeling.

\ignore{(1) {General Recommendation} with user-item interaction graphs, including NGCF~\cite{wang2019ngcf}, LightGCN~\cite{he2020lightgcn}, SGL~\cite{wu2021sgl}, HMLET~\cite{kong2022hmlet}, NCL~\cite{lin2022ncl} and SimGCL~\cite{yu2022simgcl}.
(2) {Sequential Recommendation} with item transition graphs, including SR-GNN~\cite{wu2019srgnn}, GC-SAN~\cite{xu2019gcsan}, NISER~\cite{gupta2019niser}, LESSR~\cite{chen2020lessr}, TAGNN~\cite{yu2020tagnn}, GCE-GNN~\cite{wang2020gcegnn} and SGNN-HN~\cite{pan2020sgnnhn}.
(3) {Social Recommendation} with social networks, including DiffNet~\cite{wu2019diffnet}, MHCN~\cite{yu2021mhcn} and SEPT~\cite{yu2021sept}. We also define a new category of atomic files with suffix \texttt{.net} for social networks. A series of social recommendation datasets are collected and processed into atomic files.
}

\textbf{Transformer based recommendation (RecBole-TRM).}
Another major model architecture we implement is the widely used Transformer and its variants~\cite{han2020survey}.
We implement this package by considering two major tasks, namely sequential recommendation (TiSASRec~\cite{TiSASRec}, SSE-PT~\cite{SSE-PT}, LightSANs~\cite{LightSANs}, gMLP~\cite{gMLP} and CORE~\cite{hou2022core}) and news recommendation (NRMS~\cite{NRMS}, NAML~\cite{NAML} and NPA~\cite{NPA}). 
Although the basic Transformer architectures used in these tasks are similar, the inputs of them are different, which makes Transformer play different roles in sequence modeling.
Specifically, in sequential recommendation, Transformer is utilized to capture user behavior correlations; while, in news recommendation, Transformer is utilized to extract text semantics.

\ignore{Transformer and its variants have been widely applied to building recommender models due to their strong capability for summarizing high-quality context or history information.
In order to empower our library with these latest research findings, in this package, we implement two types of models: (1) Sequential recommendation with Transformers modeling the sequence of a user's behaviors, including TiSASRec~\cite{TiSASRec}, SSE-PT~\cite{SSE-PT}, LightSANs~\cite{LightSANs} and gMLP~\cite{gMLP}. (2) News recommendation with Transformers modeling the texts and their correlations, including NRMS~\cite{NRMS}, NAML~\cite{NAML} and NPA~\cite{NPA}.
}

\textbf{Person-job recommendation (RecBole-PJF)}. Besides, 
we also include a package tailored for the task of  \emph{person-job fit}~\cite{edwards1991person}. This task is an important application and draws much attention from both research and industry communities. This package includes three categories of models, including collaborative filtering models (NeuMF~\cite{he2017neural}, LightGCN~\cite{he2020lightgcn}) and LFRR~\cite{neve2019latent}), content-based models (PJFNN~\cite{zhu2018person}, APJFNN~\cite{qin2018enhancing}, BPJFNN~\cite{qin2018enhancing} and a twin towers BERT model) and hybrid models (IPJF~\cite{le2019towards}, PJFFF~\cite{jiang2020learning} and SHPJF~\cite{hou2022leveraging}). This package designs special data mechanism to support the incorporation of additional data types. 


\noindent \emph{Package summary}. Note that each package can be run as an individual project (including the entire pipeline to produce comparison results) but with unified implementation based on RecBole. 
For clearly understanding our library, we summarize the above implemented packages and models in
Table~\ref{model}.

\begin{figure*}[t]
\centering
\setlength{\fboxrule}{0.pt}
\setlength{\fboxsep}{0.pt}
\fbox{
\includegraphics[width=.96\linewidth]{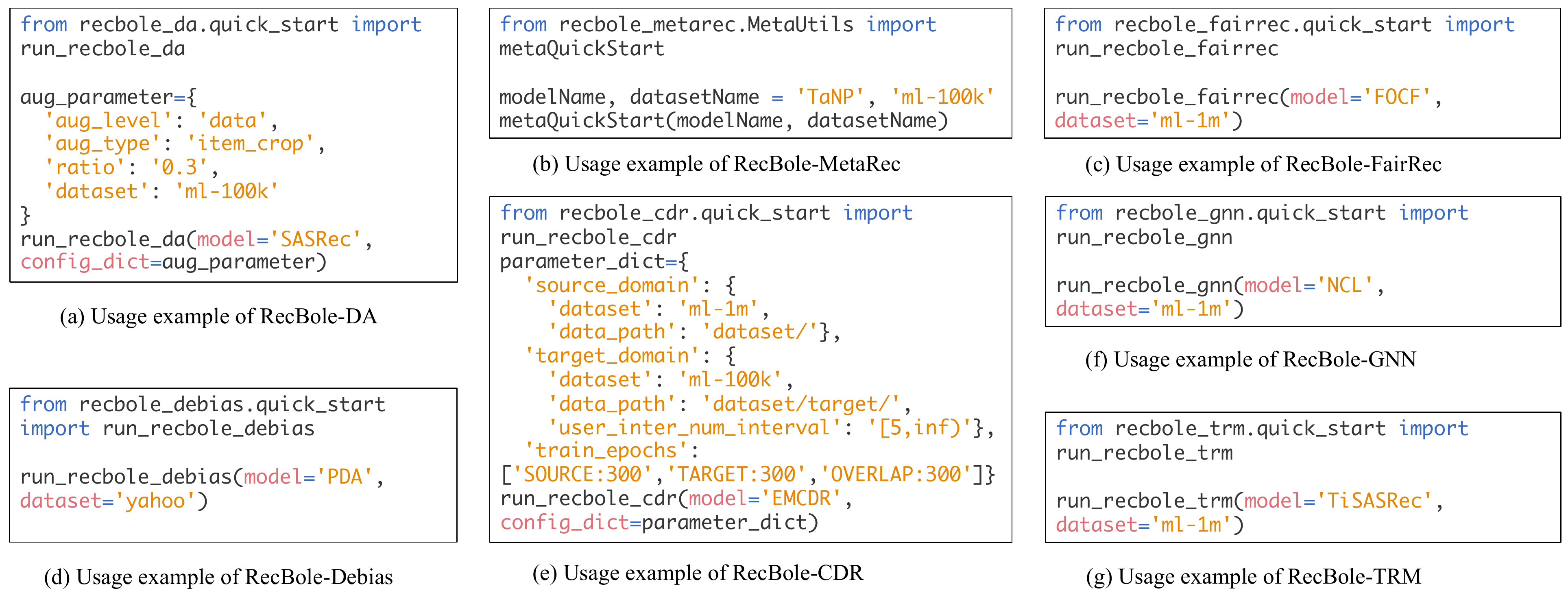}
}
\vspace{-0.1cm}
\caption{Example usage of the newly added packages in the RecBole library, where each subfigure corresponds to a package.}
\vspace{-0.1cm}
\label{fig:usage}
\end{figure*}

\vspace{-0.cm}
\section{Package Usage}
In this section, we introduce how to use our library by showing the example codes for each package separately.

\textbf{Data augmentation.}
In our library, the data augmentation package is provided through a series of augmentation interfaces or commands.
For invoking the data augmentation package, one needs to follow two steps, that is,
(1) indicating the configuration file, and (2) running the augmentation command based on existing or customized model.
Figure~\ref{fig:usage}(a) presents an example on how to run SASRec with a data augmentation strategy called ``\texttt{item\_crop}''.

\textbf{Meta recommendation.}
For running a meta learning based recommender, one can use the quick start wrapper (see Figure~\ref{fig:usage}(b)), which will automatically conduct model configuration, trainer configuration, dataset preparation, model training and model evaluation. 
To implement a new meta learning model, the users can follow three steps: 
(1) indicate the configuration file, (2) extend the \texttt{MetaRecommender} to implement the model details, and (3) extend the \texttt{MetaTrainer} to customize the training process.

\textbf{Fairness recommendation.}
For running an implemented fairness model, one should (1) indicate the parameters about the environment, data, trainer, evaluation and model by a YAML file, and (2) specify the model and dataset, and (3) launch the program with the quick-start script. 
An example to train and evaluate FOCF~\cite{FOCF}
is shown in Figure~\ref{fig:usage}(c).
In order to implement a new fairness-aware recommender model, there are three steps: 
(1) extend the \texttt{Trainer} to indicate the specific training process, (2) extend the \texttt{FairRecommender} to implement the model, and (3) extend the \texttt{AbstractMetric} to implement the fairness metrics.

\textbf{Debiased recommendation.}
To run an implemented debiased recommender model, one needs to follow two steps: (1) specify the setting of the model, dataset, training and evaluation processes via a YAML file, and (2) launch the program by indicating the model and dataset. An example of running PDA based on \textit{Yahoo!R3} is shown in Figure~\ref{fig:usage}(d). 
For implementing a model, one can implement the model architecture by extending the \texttt{DebiasedRecommender} class, and implement the trainer and sampler.

\textbf{Cross-domain recommendation.}
To run a cross-domain recommender model, the users can set the formatted dataset as either source domain or target domain by configurations and run the model with simple commands. 
We present an example of how to run EMCDR~\cite{emcdr} in our library in Figure~\ref{fig:usage}(e), where we can  set the source and target domain datasets, indicate the training modes and so on.
For implementing a new algorithm, one can firstly extend the \texttt{CrossdomainRecommender} class and then specify the training modes with simple configurations.

\textbf{GNN based recommendation.}
For running the implemented GNN models, there are two major steps to prepare: (1) indicate the customized configuration values and store them in an additional \textsc{YAML} file; (2) specify the model and dataset and launch with the quick-start script. An example to train and evaluate NCL~\cite{lin2022ncl} on \textit{Movielens-1M} dataset with customized configuration is shown in Figure~\ref{fig:usage}(f).
For implementing new GNN-based recommendation models, one can reuse or adapt GNN layers for fast reproduction. For example, we can reuse \texttt{LightGCNConv} layer in RecBole-GNN to reproduce GNN-based collaborative filtering models, or replace the graph convolutional layer to new GNN components (\emph{e.g.}, \texttt{GCN2Conv}~\cite{chen2020gcn2}) for further exploration.

\textbf{Transformer based recommendation.}
To run an implemented Transformer-based model, one can follow three steps: (1) specify the configuration via a \textsc{YAML} file; (2) indicate the dataset and model; (3) run the model with the quick-start script. An example for running TiSASRec~\cite{TiSASRec} based on \textit{Movielens-1M} dataset is shown in Figure~\ref{fig:usage}(g). 
To implement a new Transformer-based recommendation model, the users can reuse Transformer layers implemented by RecBole, or add new implementations in Transformer layers.

\textbf{Person-job recommendation}.
This package mostly follows the use of the overall RecBole library, with several adaptions or new interfaces. 
The steps for developing new models are as follows: create a new model file (\eg \texttt{PJFNN.py}), implement corresponding functions and save the hyper parameters to the configuration file.
Specially, we introduced a new parameter \emph{multi-direction} to control whether the evaluation is for biliteral  or not. If it is set to true, the evaluation will be conducted at both candidate and employer sides on the same interaction records.

Above, we have briefly introduced the usage  of the implemented packages. For more detailed descriptions, one can refer to our project at the link:  \url{https://github.com/RUCAIBox/RecBole2.0}.

\vspace{-0.cm}
\section{Discussion and Conclusions}
The included extensions are developed based on a popular recommendation library RecBole, which originally contains more than 70 recommendation models covering the tasks of general recommendation, context-aware recommendation, sequential recommendation and knowledge-based recommendation.
Since 2020, RecBole has received extension attention and use on GitHub, about 1.9K stars and 352 forks till June 1, 2022. 
In original RecBole, we focus on the design of underlying data structures, general evaluation pipeline and  classical recommendation models. 

With the rapid  progress of recommender systems, we receive increasing requests from RecBole users on the support for up-to-date advances (\eg debiased, fairness and GNNs). Meanwhile, our team members are also conducting research on these emerging topics or models. Therefore, we develop and release this extended library for enhancing RecBole by incorporating the support on recent advances of recommender systems. Specifically, in this extension, we release eight packages consisting of 65 newly implemented models, and also provide corresponding interfaces for data preparation, model running (with the well-tuned parameters) and evaluation.
We believe this extension is a significant contribution to RecBole, which is a valuable resource  to the research community. The RecBole team will continually improve this project, making it \emph{up-to-date}, \emph{comprehensive}, and \emph{flexible} for research.

\bibliographystyle{ACM-Reference-Format}
\balance
\bibliography{acmart}


\begin{thebibliography}{93}


\ifx \showCODEN    \undefined \def \showCODEN     #1{\unskip}     \fi
\ifx \showDOI      \undefined \def \showDOI       #1{#1}\fi
\ifx \showISBNx    \undefined \def \showISBNx     #1{\unskip}     \fi
\ifx \showISBNxiii \undefined \def \showISBNxiii  #1{\unskip}     \fi
\ifx \showISSN     \undefined \def \showISSN      #1{\unskip}     \fi
\ifx \showLCCN     \undefined \def \showLCCN      #1{\unskip}     \fi
\ifx \shownote     \undefined \def \shownote      #1{#1}          \fi
\ifx \showarticletitle \undefined \def \showarticletitle #1{#1}   \fi
\ifx \showURL      \undefined \def \showURL       {\relax}        \fi
\providecommand\bibfield[2]{#2}
\providecommand\bibinfo[2]{#2}
\providecommand\natexlab[1]{#1}
\providecommand\showeprint[2][]{arXiv:#2}

\bibitem[Anelli et~al\mbox{.}(2021)]%
        {Elliot}
\bibfield{author}{\bibinfo{person}{Vito~Walter Anelli},
  \bibinfo{person}{Alejandro Bellog{\'{\i}}n}, \bibinfo{person}{Antonio
  Ferrara}, \bibinfo{person}{Daniele Malitesta},
  \bibinfo{person}{Felice~Antonio Merra}, \bibinfo{person}{Claudio Pomo},
  \bibinfo{person}{Francesco~Maria Donini}, {and} \bibinfo{person}{Tommaso~Di
  Noia}.} \bibinfo{year}{2021}\natexlab{}.
\newblock \showarticletitle{Elliot: {A} Comprehensive and Rigorous Framework
  for Reproducible Recommender Systems Evaluation}. In
  \bibinfo{booktitle}{\emph{{SIGIR}}}. \bibinfo{publisher}{{ACM}},
  \bibinfo{pages}{2405--2414}.
\newblock


\bibitem[Argyriou et~al\mbox{.}(2020)]%
        {microsoft_rec}
\bibfield{author}{\bibinfo{person}{Andreas Argyriou}, \bibinfo{person}{Miguel
  Gonz{\'a}lez-Fierro}, {and} \bibinfo{person}{Le Zhang}.}
  \bibinfo{year}{2020}\natexlab{}.
\newblock \showarticletitle{Microsoft Recommenders: Best Practices for
  Production-Ready Recommendation Systems}. In
  \bibinfo{booktitle}{\emph{Companion Proceedings of the Web Conference 2020}}.
  \bibinfo{pages}{50--51}.
\newblock


\bibitem[Bian et~al\mbox{.}(2021)]%
        {BianZZCHYW21}
\bibfield{author}{\bibinfo{person}{Shuqing Bian}, \bibinfo{person}{Wayne~Xin
  Zhao}, \bibinfo{person}{Kun Zhou}, \bibinfo{person}{Jing Cai},
  \bibinfo{person}{Yancheng He}, \bibinfo{person}{Cunxiang Yin}, {and}
  \bibinfo{person}{Ji{-}Rong Wen}.} \bibinfo{year}{2021}\natexlab{}.
\newblock \showarticletitle{Contrastive Curriculum Learning for Sequential User
  Behavior Modeling via Data Augmentation}. In \bibinfo{booktitle}{\emph{{CIKM}
  2021}}. \bibinfo{pages}{3737--3746}.
\newblock


\bibitem[Bin~Wu and Staniforth(2017)]%
        {neurec}
\bibfield{author}{\bibinfo{person}{Xiangnan He Xiang~Wang Bin~Wu,
  Zhongchuan~Sun} {and} \bibinfo{person}{Jonathan Staniforth}.}
  \bibinfo{year}{2017}\natexlab{}.
\newblock \showarticletitle{NeuRec}.
\newblock \bibinfo{journal}{\emph{https://github.com/wubinzzu/NeuRec}}
  (\bibinfo{year}{2017}).
\newblock


\bibitem[Bonner and Vasile(2018)]%
        {bonner2018causal}
\bibfield{author}{\bibinfo{person}{Stephen Bonner} {and}
  \bibinfo{person}{Flavian Vasile}.} \bibinfo{year}{2018}\natexlab{}.
\newblock \showarticletitle{Causal embeddings for recommendation}. In
  \bibinfo{booktitle}{\emph{Proceedings of the 12th ACM conference on
  recommender systems}}. \bibinfo{pages}{104--112}.
\newblock


\bibitem[Chen et~al\mbox{.}(2020a)]%
        {bias_survey}
\bibfield{author}{\bibinfo{person}{Jiawei Chen}, \bibinfo{person}{Hande Dong},
  \bibinfo{person}{Xiang Wang}, \bibinfo{person}{Fuli Feng},
  \bibinfo{person}{Meng Wang}, {and} \bibinfo{person}{Xiangnan He}.}
  \bibinfo{year}{2020}\natexlab{a}.
\newblock \showarticletitle{Bias and debias in recommender system: A survey and
  future directions}.
\newblock \bibinfo{journal}{\emph{arXiv preprint arXiv:2010.03240}}
  (\bibinfo{year}{2020}).
\newblock


\bibitem[Chen et~al\mbox{.}(2020b)]%
        {chen2020bias}
\bibfield{author}{\bibinfo{person}{Jiawei Chen}, \bibinfo{person}{Hande Dong},
  \bibinfo{person}{Xiang Wang}, \bibinfo{person}{Fuli Feng},
  \bibinfo{person}{Meng Wang}, {and} \bibinfo{person}{Xiangnan He}.}
  \bibinfo{year}{2020}\natexlab{b}.
\newblock \showarticletitle{Bias and debias in recommender system: A survey and
  future directions}.
\newblock \bibinfo{journal}{\emph{arXiv preprint arXiv:2010.03240}}
  (\bibinfo{year}{2020}).
\newblock


\bibitem[Chen et~al\mbox{.}(2020c)]%
        {chen2020gcn2}
\bibfield{author}{\bibinfo{person}{Ming Chen}, \bibinfo{person}{Zhewei Wei},
  \bibinfo{person}{Zengfeng Huang}, \bibinfo{person}{Bolin Ding}, {and}
  \bibinfo{person}{Yaliang Li}.} \bibinfo{year}{2020}\natexlab{c}.
\newblock \showarticletitle{Simple and Deep Graph Convolutional Networks}. In
  \bibinfo{booktitle}{\emph{{ICML}}}.
\newblock


\bibitem[Chen and Wong(2020)]%
        {chen2020lessr}
\bibfield{author}{\bibinfo{person}{Tianwen Chen} {and}
  \bibinfo{person}{Raymond~Chi{-}Wing Wong}.} \bibinfo{year}{2020}\natexlab{}.
\newblock \showarticletitle{Handling Information Loss of Graph Neural Networks
  for Session-based Recommendation}. In \bibinfo{booktitle}{\emph{{KDD}}}.
\newblock


\bibitem[da~Costa et~al\mbox{.}(2018)]%
        {case_rec}
\bibfield{author}{\bibinfo{person}{Arthur da Costa}, \bibinfo{person}{Eduardo
  Fressato}, \bibinfo{person}{Fernando Neto}, \bibinfo{person}{Marcelo
  Manzato}, {and} \bibinfo{person}{Ricardo Campello}.}
  \bibinfo{year}{2018}\natexlab{}.
\newblock \showarticletitle{Case recommender: a flexible and extensible python
  framework for recommender systems}. In \bibinfo{booktitle}{\emph{Proceedings
  of the 12th ACM Conference on Recommender Systems}}.
  \bibinfo{pages}{494--495}.
\newblock


\bibitem[Dong et~al\mbox{.}(2020)]%
        {dong2020mamo}
\bibfield{author}{\bibinfo{person}{Manqing Dong}, \bibinfo{person}{Feng Yuan},
  \bibinfo{person}{Lina Yao}, \bibinfo{person}{Xiwei Xu}, {and}
  \bibinfo{person}{Liming Zhu}.} \bibinfo{year}{2020}\natexlab{}.
\newblock \showarticletitle{Mamo: Memory-augmented meta-optimization for
  cold-start recommendation}. In \bibinfo{booktitle}{\emph{Proceedings of the
  26th ACM SIGKDD International Conference on Knowledge Discovery \& Data
  Mining}}. \bibinfo{pages}{688--697}.
\newblock


\bibitem[Edwards(1991)]%
        {edwards1991person}
\bibfield{author}{\bibinfo{person}{Jeffrey~R Edwards}.}
  \bibinfo{year}{1991}\natexlab{}.
\newblock \bibinfo{booktitle}{\emph{Person-job fit: A conceptual integration,
  literature review, and methodological critique.}}
\newblock \bibinfo{publisher}{John Wiley \& Sons}.
\newblock


\bibitem[Fan et~al\mbox{.}(2021)]%
        {LightSANs}
\bibfield{author}{\bibinfo{person}{Xinyan Fan}, \bibinfo{person}{Zheng Liu},
  \bibinfo{person}{Jianxun Lian}, \bibinfo{person}{Wayne~Xin Zhao},
  \bibinfo{person}{Xing Xie}, {and} \bibinfo{person}{Ji-Rong Wen}.}
  \bibinfo{year}{2021}\natexlab{}.
\newblock \showarticletitle{Lighter and better: low-rank decomposed
  self-attention networks for next-item recommendation}. In
  \bibinfo{booktitle}{\emph{Proceedings of the 44th International ACM SIGIR
  Conference on Research and Development in Information Retrieval}}.
  \bibinfo{pages}{1733--1737}.
\newblock


\bibitem[Finn et~al\mbox{.}(2017)]%
        {finn2017model}
\bibfield{author}{\bibinfo{person}{Chelsea Finn}, \bibinfo{person}{Pieter
  Abbeel}, {and} \bibinfo{person}{Sergey Levine}.}
  \bibinfo{year}{2017}\natexlab{}.
\newblock \showarticletitle{Model-agnostic meta-learning for fast adaptation of
  deep networks}. In \bibinfo{booktitle}{\emph{International conference on
  machine learning}}. PMLR, \bibinfo{pages}{1126--1135}.
\newblock


\bibitem[Gao et~al\mbox{.}(2019)]%
        {natr}
\bibfield{author}{\bibinfo{person}{Chen Gao}, \bibinfo{person}{Xiangning Chen},
  \bibinfo{person}{Fuli Feng}, \bibinfo{person}{Kai Zhao},
  \bibinfo{person}{Xiangnan He}, \bibinfo{person}{Yong Li}, {and}
  \bibinfo{person}{Depeng Jin}.} \bibinfo{year}{2019}\natexlab{}.
\newblock \showarticletitle{Cross-domain recommendation without sharing
  user-relevant data}. In \bibinfo{booktitle}{\emph{TheWebConf}}.
  \bibinfo{pages}{491--502}.
\newblock


\bibitem[Gao et~al\mbox{.}(2022)]%
        {gao2022kuairec}
\bibfield{author}{\bibinfo{person}{Chongming Gao}, \bibinfo{person}{Shijun Li},
  \bibinfo{person}{Wenqiang Lei}, \bibinfo{person}{Biao Li},
  \bibinfo{person}{Peng Jiang}, \bibinfo{person}{Jiawei Chen},
  \bibinfo{person}{Xiangnan He}, \bibinfo{person}{Jiaxin Mao}, {and}
  \bibinfo{person}{Tat-Seng Chua}.} \bibinfo{year}{2022}\natexlab{}.
\newblock \showarticletitle{KuaiRec: A Fully-observed Dataset for Recommender
  Systems}.
\newblock \bibinfo{journal}{\emph{arXiv preprint arXiv:2202.10842}}
  (\bibinfo{year}{2022}).
\newblock


\bibitem[Gao et~al\mbox{.}(2013)]%
        {clfm}
\bibfield{author}{\bibinfo{person}{Sheng Gao}, \bibinfo{person}{Hao Luo},
  \bibinfo{person}{Da Chen}, \bibinfo{person}{Shantao Li},
  \bibinfo{person}{Patrick Gallinari}, {and} \bibinfo{person}{Jun Guo}.}
  \bibinfo{year}{2013}\natexlab{}.
\newblock \showarticletitle{Cross-domain recommendation via cluster-level
  latent factor model}. In \bibinfo{booktitle}{\emph{Joint European conference
  on machine learning and knowledge discovery in databases}}.
\newblock


\bibitem[Ge et~al\mbox{.}(2021)]%
        {FCPO}
\bibfield{author}{\bibinfo{person}{Yingqiang Ge}, \bibinfo{person}{Shuchang
  Liu}, \bibinfo{person}{Ruoyuan Gao}, \bibinfo{person}{Yikun Xian},
  \bibinfo{person}{Yunqi Li}, \bibinfo{person}{Xiangyu Zhao},
  \bibinfo{person}{Changhua Pei}, \bibinfo{person}{Fei Sun},
  \bibinfo{person}{Junfeng Ge}, \bibinfo{person}{Wenwu Ou}, {and}
  \bibinfo{person}{Yongfeng Zhang}.} \bibinfo{year}{2021}\natexlab{}.
\newblock \showarticletitle{Towards Long-Term Fairness in Recommendation}. In
  \bibinfo{booktitle}{\emph{Proceedings of the 14th ACM International
  Conference on Web Search and Data Mining}}. \bibinfo{publisher}{Association
  for Computing Machinery}, \bibinfo{pages}{445–453}.
\newblock
\showISBNx{9781450382977}


\bibitem[Gupta et~al\mbox{.}(2019)]%
        {gupta2019niser}
\bibfield{author}{\bibinfo{person}{Priyanka Gupta}, \bibinfo{person}{Diksha
  Garg}, \bibinfo{person}{Pankaj Malhotra}, \bibinfo{person}{Lovekesh Vig},
  {and} \bibinfo{person}{Gautam~M Shroff}.} \bibinfo{year}{2019}\natexlab{}.
\newblock \showarticletitle{NISER: Normalized Item and Session Representations
  with Graph Neural Networks}.
\newblock \bibinfo{journal}{\emph{arXiv preprint arXiv:1909.04276}}
  (\bibinfo{year}{2019}).
\newblock


\bibitem[Hamilton et~al\mbox{.}(2017)]%
        {hamilton2017graphsage}
\bibfield{author}{\bibinfo{person}{William~L. Hamilton},
  \bibinfo{person}{Zhitao Ying}, {and} \bibinfo{person}{Jure Leskovec}.}
  \bibinfo{year}{2017}\natexlab{}.
\newblock \showarticletitle{Inductive Representation Learning on Large Graphs}.
  In \bibinfo{booktitle}{\emph{{NIPS}}}.
\newblock


\bibitem[Han et~al\mbox{.}(2020)]%
        {han2020survey}
\bibfield{author}{\bibinfo{person}{Kai Han}, \bibinfo{person}{Yunhe Wang},
  \bibinfo{person}{Hanting Chen}, \bibinfo{person}{Xinghao Chen},
  \bibinfo{person}{Jianyuan Guo}, \bibinfo{person}{Zhenhua Liu},
  \bibinfo{person}{Yehui Tang}, \bibinfo{person}{An Xiao},
  \bibinfo{person}{Chunjing Xu}, \bibinfo{person}{Yixing Xu}, {et~al\mbox{.}}}
  \bibinfo{year}{2020}\natexlab{}.
\newblock \showarticletitle{A survey on visual transformer}.
\newblock \bibinfo{journal}{\emph{arXiv e-prints}} (\bibinfo{year}{2020}),
  \bibinfo{pages}{arXiv--2012}.
\newblock


\bibitem[He et~al\mbox{.}(2020)]%
        {he2020lightgcn}
\bibfield{author}{\bibinfo{person}{Xiangnan He}, \bibinfo{person}{Kuan Deng},
  \bibinfo{person}{Xiang Wang}, \bibinfo{person}{Yan Li},
  \bibinfo{person}{Yongdong Zhang}, {and} \bibinfo{person}{Meng Wang}.}
  \bibinfo{year}{2020}\natexlab{}.
\newblock \showarticletitle{Lightgcn: Simplifying and powering graph
  convolution network for recommendation}. In
  \bibinfo{booktitle}{\emph{{SIGIR}}}.
\newblock


\bibitem[He et~al\mbox{.}(2017)]%
        {he2017neural}
\bibfield{author}{\bibinfo{person}{Xiangnan He}, \bibinfo{person}{Lizi Liao},
  \bibinfo{person}{Hanwang Zhang}, \bibinfo{person}{Liqiang Nie},
  \bibinfo{person}{Xia Hu}, {and} \bibinfo{person}{Tat-Seng Chua}.}
  \bibinfo{year}{2017}\natexlab{}.
\newblock \showarticletitle{Neural collaborative filtering}. In
  \bibinfo{booktitle}{\emph{{WWW}}}.
\newblock


\bibitem[Hospedales et~al\mbox{.}(2020)]%
        {hospedales2020meta}
\bibfield{author}{\bibinfo{person}{Timothy Hospedales},
  \bibinfo{person}{Antreas Antoniou}, \bibinfo{person}{Paul Micaelli}, {and}
  \bibinfo{person}{Amos Storkey}.} \bibinfo{year}{2020}\natexlab{}.
\newblock \showarticletitle{Meta-learning in neural networks: A survey}.
\newblock \bibinfo{journal}{\emph{arXiv preprint arXiv:2004.05439}}
  (\bibinfo{year}{2020}).
\newblock


\bibitem[Hou et~al\mbox{.}(2022a)]%
        {hou2022core}
\bibfield{author}{\bibinfo{person}{Yupeng Hou}, \bibinfo{person}{Binbin Hu},
  \bibinfo{person}{Zhiqiang Zhang}, {and} \bibinfo{person}{Wayne~Xin Zhao}.}
  \bibinfo{year}{2022}\natexlab{a}.
\newblock \showarticletitle{CORE: Simple and Effective Session-based
  Recommendation within Consistent Representation Space}. In
  \bibinfo{booktitle}{\emph{{SIGIR}}}.
\newblock


\bibitem[Hou et~al\mbox{.}(2022b)]%
        {hou2022leveraging}
\bibfield{author}{\bibinfo{person}{Yupeng Hou}, \bibinfo{person}{Xingyu Pan},
  \bibinfo{person}{Wayne~Xin Zhao}, \bibinfo{person}{Shuqing Bian},
  \bibinfo{person}{Yang Song}, \bibinfo{person}{Tao Zhang}, {and}
  \bibinfo{person}{Ji-Rong Wen}.} \bibinfo{year}{2022}\natexlab{b}.
\newblock \showarticletitle{Leveraging Search History for Improving Person-Job
  Fit}. In \bibinfo{booktitle}{\emph{Database Systems for Advanced
  Applications: 27th International Conference, DASFAA 2022, Virtual Event,
  April 11--14, 2022, Proceedings, Part I}}. \bibinfo{pages}{38--54}.
\newblock


\bibitem[Hu et~al\mbox{.}(2018)]%
        {conet}
\bibfield{author}{\bibinfo{person}{Guangneng Hu}, \bibinfo{person}{Yu Zhang},
  {and} \bibinfo{person}{Qiang Yang}.} \bibinfo{year}{2018}\natexlab{}.
\newblock \showarticletitle{Conet: Collaborative cross networks for
  cross-domain recommendation}. In \bibinfo{booktitle}{\emph{CIKM}}.
\newblock


\bibitem[Hug(2020)]%
        {surprise}
\bibfield{author}{\bibinfo{person}{Nicolas Hug}.}
  \bibinfo{year}{2020}\natexlab{}.
\newblock \showarticletitle{Surprise: A Python library for recommender
  systems}.
\newblock \bibinfo{journal}{\emph{Journal of Open Source Software}}
  \bibinfo{volume}{5}, \bibinfo{number}{52} (\bibinfo{year}{2020}),
  \bibinfo{pages}{2174}.
\newblock


\bibitem[Huisman et~al\mbox{.}(2021)]%
        {huisman2021survey}
\bibfield{author}{\bibinfo{person}{Mike Huisman}, \bibinfo{person}{Jan~N
  Van~Rijn}, {and} \bibinfo{person}{Aske Plaat}.}
  \bibinfo{year}{2021}\natexlab{}.
\newblock \showarticletitle{A survey of deep meta-learning}.
\newblock \bibinfo{journal}{\emph{Artificial Intelligence Review}}
  \bibinfo{volume}{54}, \bibinfo{number}{6} (\bibinfo{year}{2021}),
  \bibinfo{pages}{4483--4541}.
\newblock


\bibitem[Islam et~al\mbox{.}(2021)]%
        {NFCF}
\bibfield{author}{\bibinfo{person}{Rashidul Islam},
  \bibinfo{person}{Kamrun~Naher Keya}, \bibinfo{person}{Ziqian Zeng},
  \bibinfo{person}{Shimei Pan}, {and} \bibinfo{person}{James Foulds}.}
  \bibinfo{year}{2021}\natexlab{}.
\newblock \showarticletitle{Debiasing Career Recommendations with Neural Fair
  Collaborative Filtering}. In \bibinfo{booktitle}{\emph{Proceedings of the Web
  Conference 2021}}. \bibinfo{pages}{3779–3790}.
\newblock
\showISBNx{9781450383127}


\bibitem[Jiang et~al\mbox{.}(2020)]%
        {jiang2020learning}
\bibfield{author}{\bibinfo{person}{Junshu Jiang}, \bibinfo{person}{Songyun Ye},
  \bibinfo{person}{Wei Wang}, \bibinfo{person}{Jingran Xu}, {and}
  \bibinfo{person}{Xiaosheng Luo}.} \bibinfo{year}{2020}\natexlab{}.
\newblock \showarticletitle{Learning Effective Representations for Person-Job
  Fit by Feature Fusion}. In \bibinfo{booktitle}{\emph{{CIKM}}}.
\newblock


\bibitem[Kamishima et~al\mbox{.}(2011)]%
        {NonParity}
\bibfield{author}{\bibinfo{person}{Toshihiro Kamishima},
  \bibinfo{person}{Shotaro Akaho}, {and} \bibinfo{person}{Jun Sakuma}.}
  \bibinfo{year}{2011}\natexlab{}.
\newblock \showarticletitle{Fairness-aware Learning through Regularization
  Approach}. In \bibinfo{booktitle}{\emph{2011 IEEE 11th International
  Conference on Data Mining Workshops}}. \bibinfo{pages}{643--650}.
\newblock


\bibitem[Kang et~al\mbox{.}(2019)]%
        {sscdr}
\bibfield{author}{\bibinfo{person}{SeongKu Kang}, \bibinfo{person}{Junyoung
  Hwang}, \bibinfo{person}{Dongha Lee}, {and} \bibinfo{person}{Hwanjo Yu}.}
  \bibinfo{year}{2019}\natexlab{}.
\newblock \showarticletitle{Semi-supervised learning for cross-domain
  recommendation to cold-start users}. In \bibinfo{booktitle}{\emph{CIKM}}.
\newblock


\bibitem[Kipf and Welling(2017)]%
        {kipf2017gcn}
\bibfield{author}{\bibinfo{person}{Thomas~N. Kipf} {and} \bibinfo{person}{Max
  Welling}.} \bibinfo{year}{2017}\natexlab{}.
\newblock \showarticletitle{Semi-Supervised Classification with Graph
  Convolutional Networks}. In \bibinfo{booktitle}{\emph{{ICLR}}}.
\newblock


\bibitem[Kong et~al\mbox{.}(2022)]%
        {kong2022hmlet}
\bibfield{author}{\bibinfo{person}{Taeyong Kong}, \bibinfo{person}{Taeri Kim},
  \bibinfo{person}{Jinsung Jeon}, \bibinfo{person}{Jeongwhan Choi},
  \bibinfo{person}{Yeon{-}Chang Lee}, \bibinfo{person}{Noseong Park}, {and}
  \bibinfo{person}{Sang{-}Wook Kim}.} \bibinfo{year}{2022}\natexlab{}.
\newblock \showarticletitle{Linear, or Non-Linear, That is the Question!}. In
  \bibinfo{booktitle}{\emph{{WSDM}}}.
\newblock


\bibitem[Le et~al\mbox{.}(2019)]%
        {le2019towards}
\bibfield{author}{\bibinfo{person}{Ran Le}, \bibinfo{person}{Wenpeng Hu},
  \bibinfo{person}{Yang Song}, \bibinfo{person}{Tao Zhang},
  \bibinfo{person}{Dongyan Zhao}, {and} \bibinfo{person}{Rui Yan}.}
  \bibinfo{year}{2019}\natexlab{}.
\newblock \showarticletitle{Towards effective and interpretable person-job
  fitting}. In \bibinfo{booktitle}{\emph{{CIKM}}}.
\newblock


\bibitem[Lee et~al\mbox{.}(2019)]%
        {lee2019melu}
\bibfield{author}{\bibinfo{person}{Hoyeop Lee}, \bibinfo{person}{Jinbae Im},
  \bibinfo{person}{Seongwon Jang}, \bibinfo{person}{Hyunsouk Cho}, {and}
  \bibinfo{person}{Sehee Chung}.} \bibinfo{year}{2019}\natexlab{}.
\newblock \showarticletitle{Melu: Meta-learned user preference estimator for
  cold-start recommendation}. In \bibinfo{booktitle}{\emph{Proceedings of the
  25th ACM SIGKDD International Conference on Knowledge Discovery \& Data
  Mining}}. \bibinfo{pages}{1073--1082}.
\newblock


\bibitem[Li et~al\mbox{.}(2020)]%
        {TiSASRec}
\bibfield{author}{\bibinfo{person}{Jiacheng Li}, \bibinfo{person}{Yujie Wang},
  {and} \bibinfo{person}{Julian~J. McAuley}.} \bibinfo{year}{2020}\natexlab{}.
\newblock \showarticletitle{Time Interval Aware Self-Attention for Sequential
  Recommendation}. In \bibinfo{booktitle}{\emph{{WSDM} '20: The Thirteenth
  {ACM} International Conference on Web Search and Data Mining, Houston, TX,
  USA, February 3-7, 2020}}, \bibfield{editor}{\bibinfo{person}{James
  Caverlee}, \bibinfo{person}{Xia~(Ben) Hu}, \bibinfo{person}{Mounia Lalmas},
  {and} \bibinfo{person}{Wei Wang}} (Eds.). \bibinfo{pages}{322--330}.
\newblock


\bibitem[Li et~al\mbox{.}(2021)]%
        {PFCN}
\bibfield{author}{\bibinfo{person}{Yunqi Li}, \bibinfo{person}{Hanxiong Chen},
  \bibinfo{person}{Shuyuan Xu}, \bibinfo{person}{Yingqiang Ge}, {and}
  \bibinfo{person}{Yongfeng Zhang}.} \bibinfo{year}{2021}\natexlab{}.
\newblock \bibinfo{booktitle}{\emph{Towards Personalized Fairness Based on
  Causal Notion}}.
\newblock \bibinfo{publisher}{Association for Computing Machinery},
  \bibinfo{pages}{1054–1063}.
\newblock
\showISBNx{9781450380379}


\bibitem[Lin et~al\mbox{.}(2021)]%
        {lin2021task}
\bibfield{author}{\bibinfo{person}{Xixun Lin}, \bibinfo{person}{Jia Wu},
  \bibinfo{person}{Chuan Zhou}, \bibinfo{person}{Shirui Pan},
  \bibinfo{person}{Yanan Cao}, {and} \bibinfo{person}{Bin Wang}.}
  \bibinfo{year}{2021}\natexlab{}.
\newblock \showarticletitle{Task-adaptive neural process for user cold-start
  recommendation}. In \bibinfo{booktitle}{\emph{Proceedings of the Web
  Conference 2021}}. \bibinfo{pages}{1306--1316}.
\newblock


\bibitem[Lin et~al\mbox{.}(2022)]%
        {lin2022ncl}
\bibfield{author}{\bibinfo{person}{Zihan Lin}, \bibinfo{person}{Changxin Tian},
  \bibinfo{person}{Yupeng Hou}, {and} \bibinfo{person}{Wayne~Xin Zhao}.}
  \bibinfo{year}{2022}\natexlab{}.
\newblock \showarticletitle{Improving Graph Collaborative Filtering with
  Neighborhood-enriched Contrastive Learning}. In
  \bibinfo{booktitle}{\emph{{TheWebConf}}}.
\newblock


\bibitem[Liu et~al\mbox{.}(2021b)]%
        {gMLP}
\bibfield{author}{\bibinfo{person}{Hanxiao Liu}, \bibinfo{person}{Zihang Dai},
  \bibinfo{person}{David~R So}, {and} \bibinfo{person}{Quoc~V Le}.}
  \bibinfo{year}{2021}\natexlab{b}.
\newblock \showarticletitle{Pay Attention to MLPs}.
\newblock \bibinfo{journal}{\emph{arXiv preprint arXiv:2105.08050}}
  (\bibinfo{year}{2021}).
\newblock


\bibitem[Liu et~al\mbox{.}(2020)]%
        {bitgcf}
\bibfield{author}{\bibinfo{person}{Meng Liu}, \bibinfo{person}{Jianjun Li},
  \bibinfo{person}{Guohui Li}, {and} \bibinfo{person}{Peng Pan}.}
  \bibinfo{year}{2020}\natexlab{}.
\newblock \showarticletitle{Cross domain recommendation via bi-directional
  transfer graph collaborative filtering networks}. In
  \bibinfo{booktitle}{\emph{CIKM}}.
\newblock


\bibitem[Liu et~al\mbox{.}(2021a)]%
        {abs-2108-06479}
\bibfield{author}{\bibinfo{person}{Zhiwei Liu}, \bibinfo{person}{Yongjun Chen},
  \bibinfo{person}{Jia Li}, \bibinfo{person}{Philip~S. Yu},
  \bibinfo{person}{Julian~J. McAuley}, {and} \bibinfo{person}{Caiming Xiong}.}
  \bibinfo{year}{2021}\natexlab{a}.
\newblock \showarticletitle{Contrastive Self-supervised Sequential
  Recommendation with Robust Augmentation}.
\newblock \bibinfo{journal}{\emph{CoRR}}  \bibinfo{volume}{abs/2108.06479}
  (\bibinfo{year}{2021}).
\newblock


\bibitem[Man et~al\mbox{.}(2017)]%
        {emcdr}
\bibfield{author}{\bibinfo{person}{Tong Man}, \bibinfo{person}{Huawei Shen},
  \bibinfo{person}{Xiaolong Jin}, {and} \bibinfo{person}{Xueqi Cheng}.}
  \bibinfo{year}{2017}\natexlab{}.
\newblock \showarticletitle{Cross-domain recommendation: An embedding and
  mapping approach.}. In \bibinfo{booktitle}{\emph{IJCAI}}.
\newblock


\bibitem[Meng et~al\mbox{.}(2020)]%
        {beta-rec}
\bibfield{author}{\bibinfo{person}{Zaiqiao Meng}, \bibinfo{person}{Richard
  McCreadie}, \bibinfo{person}{Craig Macdonald}, \bibinfo{person}{Iadh Ounis},
  \bibinfo{person}{Siwei Liu}, \bibinfo{person}{Yaxiong Wu},
  \bibinfo{person}{Xi Wang}, \bibinfo{person}{Shangsong Liang},
  \bibinfo{person}{Yucheng Liang}, \bibinfo{person}{Guangtao Zeng},
  {et~al\mbox{.}}} \bibinfo{year}{2020}\natexlab{}.
\newblock \showarticletitle{Beta-rec: Build, evaluate and tune automated
  recommender systems}. In \bibinfo{booktitle}{\emph{Fourteenth ACM conference
  on recommender systems}}. \bibinfo{pages}{588--590}.
\newblock


\bibitem[Neve and Palomares(2019)]%
        {neve2019latent}
\bibfield{author}{\bibinfo{person}{James Neve} {and} \bibinfo{person}{Ivan
  Palomares}.} \bibinfo{year}{2019}\natexlab{}.
\newblock \showarticletitle{Latent factor models and aggregation operators for
  collaborative filtering in reciprocal recommender systems}. In
  \bibinfo{booktitle}{\emph{{RecSys}}}.
\newblock


\bibitem[Nichol et~al\mbox{.}(2018)]%
        {nichol2018first}
\bibfield{author}{\bibinfo{person}{Alex Nichol}, \bibinfo{person}{Joshua
  Achiam}, {and} \bibinfo{person}{John Schulman}.}
  \bibinfo{year}{2018}\natexlab{}.
\newblock \showarticletitle{On first-order meta-learning algorithms}.
\newblock \bibinfo{journal}{\emph{arXiv preprint arXiv:1803.02999}}
  (\bibinfo{year}{2018}).
\newblock


\bibitem[Pan et~al\mbox{.}(2019)]%
        {pan2019warm}
\bibfield{author}{\bibinfo{person}{Feiyang Pan}, \bibinfo{person}{Shuokai Li},
  \bibinfo{person}{Xiang Ao}, \bibinfo{person}{Pingzhong Tang}, {and}
  \bibinfo{person}{Qing He}.} \bibinfo{year}{2019}\natexlab{}.
\newblock \showarticletitle{Warm up cold-start advertisements: Improving ctr
  predictions via learning to learn id embeddings}. In
  \bibinfo{booktitle}{\emph{Proceedings of the 42nd International ACM SIGIR
  Conference on Research and Development in Information Retrieval}}.
  \bibinfo{pages}{695--704}.
\newblock


\bibitem[Pan et~al\mbox{.}(2020)]%
        {pan2020sgnnhn}
\bibfield{author}{\bibinfo{person}{Zhiqiang Pan}, \bibinfo{person}{Fei Cai},
  \bibinfo{person}{Wanyu Chen}, \bibinfo{person}{Honghui Chen}, {and}
  \bibinfo{person}{Maarten de Rijke}.} \bibinfo{year}{2020}\natexlab{}.
\newblock \showarticletitle{Star Graph Neural Networks for Session-based
  Recommendation}. In \bibinfo{booktitle}{\emph{{CIKM}}}.
\newblock


\bibitem[Pitoura et~al\mbox{.}(2022)]%
        {fairness_survey}
\bibfield{author}{\bibinfo{person}{Evaggelia Pitoura}, \bibinfo{person}{Kostas
  Stefanidis}, {and} \bibinfo{person}{Georgia Koutrika}.}
  \bibinfo{year}{2022}\natexlab{}.
\newblock \showarticletitle{Fairness in rankings and recommendations: an
  overview}.
\newblock \bibinfo{journal}{\emph{{VLDB} J.}} \bibinfo{volume}{31},
  \bibinfo{number}{3} (\bibinfo{year}{2022}), \bibinfo{pages}{431--458}.
\newblock


\bibitem[Qin et~al\mbox{.}(2018)]%
        {qin2018enhancing}
\bibfield{author}{\bibinfo{person}{Chuan Qin}, \bibinfo{person}{Hengshu Zhu},
  \bibinfo{person}{Tong Xu}, \bibinfo{person}{Chen Zhu}, \bibinfo{person}{Liang
  Jiang}, \bibinfo{person}{Enhong Chen}, {and} \bibinfo{person}{Hui Xiong}.}
  \bibinfo{year}{2018}\natexlab{}.
\newblock \showarticletitle{Enhancing person-job fit for talent recruitment: An
  ability-aware neural network approach}. In
  \bibinfo{booktitle}{\emph{{SIGIR}}}.
\newblock


\bibitem[Qiu et~al\mbox{.}(2021)]%
        {QiuHY21}
\bibfield{author}{\bibinfo{person}{Ruihong Qiu}, \bibinfo{person}{Zi Huang},
  {and} \bibinfo{person}{Hongzhi Yin}.} \bibinfo{year}{2021}\natexlab{}.
\newblock \showarticletitle{Memory Augmented Multi-Instance Contrastive
  Predictive Coding for Sequential Recommendation}. In
  \bibinfo{booktitle}{\emph{{ICDM} 2021}}. \bibinfo{pages}{519--528}.
\newblock


\bibitem[Qiu et~al\mbox{.}(2022)]%
        {abs-2110-05730}
\bibfield{author}{\bibinfo{person}{Ruihong Qiu}, \bibinfo{person}{Zi Huang},
  \bibinfo{person}{Hongzhi Yin}, {and} \bibinfo{person}{Zijian Wang}.}
  \bibinfo{year}{2022}\natexlab{}.
\newblock \showarticletitle{Contrastive Learning for Representation
  Degeneration Problem in Sequential Recommendation}. In
  \bibinfo{booktitle}{\emph{{WSDM} 2022}}.
\newblock


\bibitem[Saito et~al\mbox{.}(2020)]%
        {saito2020unbiased}
\bibfield{author}{\bibinfo{person}{Yuta Saito}, \bibinfo{person}{Suguru
  Yaginuma}, \bibinfo{person}{Yuta Nishino}, \bibinfo{person}{Hayato Sakata},
  {and} \bibinfo{person}{Kazuhide Nakata}.} \bibinfo{year}{2020}\natexlab{}.
\newblock \showarticletitle{Unbiased recommender learning from
  missing-not-at-random implicit feedback}. In
  \bibinfo{booktitle}{\emph{Proceedings of the 13th International Conference on
  Web Search and Data Mining}}. \bibinfo{pages}{501--509}.
\newblock


\bibitem[Schnabel et~al\mbox{.}(2016)]%
        {schnabel2016recommendations}
\bibfield{author}{\bibinfo{person}{Tobias Schnabel}, \bibinfo{person}{Adith
  Swaminathan}, \bibinfo{person}{Ashudeep Singh}, \bibinfo{person}{Navin
  Chandak}, {and} \bibinfo{person}{Thorsten Joachims}.}
  \bibinfo{year}{2016}\natexlab{}.
\newblock \showarticletitle{Recommendations as treatments: Debiasing learning
  and evaluation}. In \bibinfo{booktitle}{\emph{international conference on
  machine learning}}. PMLR, \bibinfo{pages}{1670--1679}.
\newblock


\bibitem[Singh and Gordon(2008)]%
        {cmf}
\bibfield{author}{\bibinfo{person}{Ajit~P Singh} {and}
  \bibinfo{person}{Geoffrey~J Gordon}.} \bibinfo{year}{2008}\natexlab{}.
\newblock \showarticletitle{Relational learning via collective matrix
  factorization}. In \bibinfo{booktitle}{\emph{Proceedings of the 14th ACM
  SIGKDD international conference on Knowledge discovery and data mining}}.
  \bibinfo{pages}{650--658}.
\newblock


\bibitem[Sun et~al\mbox{.}(2020)]%
        {DaisyRec}
\bibfield{author}{\bibinfo{person}{Zhu Sun}, \bibinfo{person}{Di Yu},
  \bibinfo{person}{Hui Fang}, \bibinfo{person}{Jie Yang},
  \bibinfo{person}{Xinghua Qu}, \bibinfo{person}{Jie Zhang}, {and}
  \bibinfo{person}{Cong Geng}.} \bibinfo{year}{2020}\natexlab{}.
\newblock \showarticletitle{Are We Evaluating Rigorously? Benchmarking
  Recommendation for Reproducible Evaluation and Fair Comparison}. In
  \bibinfo{booktitle}{\emph{RecSys}}. \bibinfo{publisher}{{ACM}},
  \bibinfo{pages}{23--32}.
\newblock


\bibitem[Vartak et~al\mbox{.}(2017)]%
        {vartak2017meta}
\bibfield{author}{\bibinfo{person}{Manasi Vartak}, \bibinfo{person}{Arvind
  Thiagarajan}, \bibinfo{person}{Conrado Miranda}, \bibinfo{person}{Jeshua
  Bratman}, {and} \bibinfo{person}{Hugo Larochelle}.}
  \bibinfo{year}{2017}\natexlab{}.
\newblock \showarticletitle{A meta-learning perspective on cold-start
  recommendations for items}.
\newblock \bibinfo{journal}{\emph{Advances in neural information processing
  systems}}  \bibinfo{volume}{30} (\bibinfo{year}{2017}).
\newblock


\bibitem[Velickovic et~al\mbox{.}(2018)]%
        {velickovic2018gat}
\bibfield{author}{\bibinfo{person}{Petar Velickovic}, \bibinfo{person}{Guillem
  Cucurull}, \bibinfo{person}{Arantxa Casanova}, \bibinfo{person}{Adriana
  Romero}, \bibinfo{person}{Pietro Li{\`{o}}}, {and} \bibinfo{person}{Yoshua
  Bengio}.} \bibinfo{year}{2018}\natexlab{}.
\newblock \showarticletitle{Graph Attention Networks}. In
  \bibinfo{booktitle}{\emph{{ICLR}}}.
\newblock


\bibitem[Wang et~al\mbox{.}(2020b)]%
        {Chorus}
\bibfield{author}{\bibinfo{person}{Chenyang Wang}, \bibinfo{person}{Min Zhang},
  \bibinfo{person}{Weizhi Ma}, \bibinfo{person}{Yiqun Liu}, {and}
  \bibinfo{person}{Shaoping Ma}.} \bibinfo{year}{2020}\natexlab{b}.
\newblock \showarticletitle{Make It a Chorus: Knowledge- and Time-aware Item
  Modeling for Sequential Recommendation}. In
  \bibinfo{booktitle}{\emph{{SIGIR}}}. \bibinfo{publisher}{{ACM}},
  \bibinfo{pages}{109--118}.
\newblock


\bibitem[Wang et~al\mbox{.}(2019)]%
        {wang2019ngcf}
\bibfield{author}{\bibinfo{person}{Xiang Wang}, \bibinfo{person}{Xiangnan He},
  \bibinfo{person}{Meng Wang}, \bibinfo{person}{Fuli Feng}, {and}
  \bibinfo{person}{Tat-Seng Chua}.} \bibinfo{year}{2019}\natexlab{}.
\newblock \showarticletitle{Neural graph collaborative filtering}. In
  \bibinfo{booktitle}{\emph{{SIGIR}}}.
\newblock


\bibitem[Wang et~al\mbox{.}(2020a)]%
        {wang2020gcegnn}
\bibfield{author}{\bibinfo{person}{Ziyang Wang}, \bibinfo{person}{Wei Wei},
  \bibinfo{person}{Gao Cong}, \bibinfo{person}{Xiao{-}Li Li},
  \bibinfo{person}{Xianling Mao}, {and} \bibinfo{person}{Minghui Qiu}.}
  \bibinfo{year}{2020}\natexlab{a}.
\newblock \showarticletitle{Global Context Enhanced Graph Neural Networks for
  Session-based Recommendation}. In \bibinfo{booktitle}{\emph{{SIGIR}}}.
\newblock


\bibitem[Wang et~al\mbox{.}(2021)]%
        {WangZXCZZW21}
\bibfield{author}{\bibinfo{person}{Zhenlei Wang}, \bibinfo{person}{Jingsen
  Zhang}, \bibinfo{person}{Hongteng Xu}, \bibinfo{person}{Xu Chen},
  \bibinfo{person}{Yongfeng Zhang}, \bibinfo{person}{Wayne~Xin Zhao}, {and}
  \bibinfo{person}{Ji{-}Rong Wen}.} \bibinfo{year}{2021}\natexlab{}.
\newblock \showarticletitle{Counterfactual Data-Augmented Sequential
  Recommendation}. In \bibinfo{booktitle}{\emph{{SIGIR} 2021}}.
  \bibinfo{pages}{347--356}.
\newblock


\bibitem[Wei et~al\mbox{.}(2021)]%
        {wei2021model}
\bibfield{author}{\bibinfo{person}{Tianxin Wei}, \bibinfo{person}{Fuli Feng},
  \bibinfo{person}{Jiawei Chen}, \bibinfo{person}{Ziwei Wu},
  \bibinfo{person}{Jinfeng Yi}, {and} \bibinfo{person}{Xiangnan He}.}
  \bibinfo{year}{2021}\natexlab{}.
\newblock \showarticletitle{Model-agnostic counterfactual reasoning for
  eliminating popularity bias in recommender system}. In
  \bibinfo{booktitle}{\emph{Proceedings of the 27th ACM SIGKDD Conference on
  Knowledge Discovery \& Data Mining}}. \bibinfo{pages}{1791--1800}.
\newblock


\bibitem[Wu et~al\mbox{.}(2019c)]%
        {NAML}
\bibfield{author}{\bibinfo{person}{Chuhan Wu}, \bibinfo{person}{Fangzhao Wu},
  \bibinfo{person}{Mingxiao An}, \bibinfo{person}{Jianqiang Huang},
  \bibinfo{person}{Yongfeng Huang}, {and} \bibinfo{person}{Xing Xie}.}
  \bibinfo{year}{2019}\natexlab{c}.
\newblock \showarticletitle{Neural News Recommendation with Attentive
  Multi-View Learning}. In \bibinfo{booktitle}{\emph{Proceedings of the
  Twenty-Eighth International Joint Conference on Artificial Intelligence,
  {IJCAI} 2019, Macao, China, August 10-16, 2019}}.
  \bibinfo{pages}{3863--3869}.
\newblock


\bibitem[Wu et~al\mbox{.}(2019d)]%
        {NPA}
\bibfield{author}{\bibinfo{person}{Chuhan Wu}, \bibinfo{person}{Fangzhao Wu},
  \bibinfo{person}{Mingxiao An}, \bibinfo{person}{Jianqiang Huang},
  \bibinfo{person}{Yongfeng Huang}, {and} \bibinfo{person}{Xing Xie}.}
  \bibinfo{year}{2019}\natexlab{d}.
\newblock \showarticletitle{{NPA:} Neural News Recommendation with Personalized
  Attention}. In \bibinfo{booktitle}{\emph{Proceedings of the 25th {ACM}
  {SIGKDD} International Conference on Knowledge Discovery {\&} Data Mining,
  {KDD} 2019, Anchorage, AK, USA, August 4-8, 2019}}.
  \bibinfo{pages}{2576--2584}.
\newblock


\bibitem[Wu et~al\mbox{.}(2019e)]%
        {NRMS}
\bibfield{author}{\bibinfo{person}{Chuhan Wu}, \bibinfo{person}{Fangzhao Wu},
  \bibinfo{person}{Suyu Ge}, \bibinfo{person}{Tao Qi},
  \bibinfo{person}{Yongfeng Huang}, {and} \bibinfo{person}{Xing Xie}.}
  \bibinfo{year}{2019}\natexlab{e}.
\newblock \showarticletitle{Neural News Recommendation with Multi-Head
  Self-Attention}. In \bibinfo{booktitle}{\emph{Proceedings of the 2019
  Conference on Empirical Methods in Natural Language Processing and the 9th
  International Joint Conference on Natural Language Processing, {EMNLP-IJCNLP}
  2019, Hong Kong, China, November 3-7, 2019}}. \bibinfo{pages}{6388--6393}.
\newblock


\bibitem[Wu et~al\mbox{.}(2021b)]%
        {wu2021sgl}
\bibfield{author}{\bibinfo{person}{Jiancan Wu}, \bibinfo{person}{Xiang Wang},
  \bibinfo{person}{Fuli Feng}, \bibinfo{person}{Xiangnan He},
  \bibinfo{person}{Liang Chen}, \bibinfo{person}{Jianxun Lian}, {and}
  \bibinfo{person}{Xing Xie}.} \bibinfo{year}{2021}\natexlab{b}.
\newblock \showarticletitle{Self-supervised graph learning for recommendation}.
  In \bibinfo{booktitle}{\emph{{SIGIR}}}.
\newblock


\bibitem[Wu et~al\mbox{.}(2021a)]%
        {FairGo}
\bibfield{author}{\bibinfo{person}{Le Wu}, \bibinfo{person}{Lei Chen},
  \bibinfo{person}{Pengyang Shao}, \bibinfo{person}{Richang Hong},
  \bibinfo{person}{Xiting Wang}, {and} \bibinfo{person}{Meng Wang}.}
  \bibinfo{year}{2021}\natexlab{a}.
\newblock \showarticletitle{Learning Fair Representations for Recommendation: A
  Graph-Based Perspective}. In \bibinfo{booktitle}{\emph{Proceedings of the Web
  Conference 2021}}. \bibinfo{publisher}{Association for Computing Machinery},
  \bibinfo{address}{New York, NY, USA}, \bibinfo{pages}{2198–2208}.
\newblock
\showISBNx{9781450383127}


\bibitem[Wu et~al\mbox{.}(2020)]%
        {SSE-PT}
\bibfield{author}{\bibinfo{person}{Liwei Wu}, \bibinfo{person}{Shuqing Li},
  \bibinfo{person}{Cho{-}Jui Hsieh}, {and} \bibinfo{person}{James Sharpnack}.}
  \bibinfo{year}{2020}\natexlab{}.
\newblock \showarticletitle{{SSE-PT:} Sequential Recommendation Via
  Personalized Transformer}. In \bibinfo{booktitle}{\emph{RecSys 2020:
  Fourteenth {ACM} Conference on Recommender Systems, Virtual Event, Brazil,
  September 22-26, 2020}}. \bibinfo{pages}{328--337}.
\newblock


\bibitem[Wu et~al\mbox{.}(2019a)]%
        {wu2019diffnet}
\bibfield{author}{\bibinfo{person}{Le Wu}, \bibinfo{person}{Peijie Sun},
  \bibinfo{person}{Yanjie Fu}, \bibinfo{person}{Richang Hong},
  \bibinfo{person}{Xiting Wang}, {and} \bibinfo{person}{Meng Wang}.}
  \bibinfo{year}{2019}\natexlab{a}.
\newblock \showarticletitle{A Neural Influence Diffusion Model for Social
  Recommendation}. In \bibinfo{booktitle}{\emph{{SIGIR}}}.
\newblock


\bibitem[Wu et~al\mbox{.}(2022)]%
        {wu2022gnn4rec_survey}
\bibfield{author}{\bibinfo{person}{Shiwen Wu}, \bibinfo{person}{Fei Sun},
  \bibinfo{person}{Wentao Zhang}, \bibinfo{person}{Xu Xie}, {and}
  \bibinfo{person}{Bin Cui}.} \bibinfo{year}{2022}\natexlab{}.
\newblock \showarticletitle{Graph Neural Networks in Recommender Systems: A
  Survey}.
\newblock \bibinfo{journal}{\emph{ACM Comput. Surv.}} (\bibinfo{year}{2022}).
\newblock


\bibitem[Wu et~al\mbox{.}(2019b)]%
        {wu2019srgnn}
\bibfield{author}{\bibinfo{person}{Shu Wu}, \bibinfo{person}{Yuyuan Tang},
  \bibinfo{person}{Yanqiao Zhu}, \bibinfo{person}{Liang Wang},
  \bibinfo{person}{Xing Xie}, {and} \bibinfo{person}{Tieniu Tan}.}
  \bibinfo{year}{2019}\natexlab{b}.
\newblock \showarticletitle{Session-Based Recommendation with Graph Neural
  Networks}. In \bibinfo{booktitle}{\emph{{AAAI}}}.
\newblock


\bibitem[Xie et~al\mbox{.}(2020)]%
        {DBLP:journals/corr/abs-2010-14395}
\bibfield{author}{\bibinfo{person}{Xu Xie}, \bibinfo{person}{Fei Sun},
  \bibinfo{person}{Bolin Ding}, {and} \bibinfo{person}{Bin Cui}.}
  \bibinfo{year}{2020}\natexlab{}.
\newblock \showarticletitle{Contrastive Pre-training for Sequential
  Recommendation}.
\newblock \bibinfo{journal}{\emph{CoRR}}  \bibinfo{volume}{abs/2010.14395}
  (\bibinfo{year}{2020}).
\newblock


\bibitem[Xu et~al\mbox{.}(2019)]%
        {xu2019gcsan}
\bibfield{author}{\bibinfo{person}{Chengfeng Xu}, \bibinfo{person}{Pengpeng
  Zhao}, \bibinfo{person}{Yanchi Liu}, \bibinfo{person}{Victor~S. Sheng},
  \bibinfo{person}{Jiajie Xu}, \bibinfo{person}{Fuzhen Zhuang},
  \bibinfo{person}{Junhua Fang}, {and} \bibinfo{person}{Xiaofang Zhou}.}
  \bibinfo{year}{2019}\natexlab{}.
\newblock \showarticletitle{Graph Contextualized Self-Attention Network for
  Session-based Recommendation}. In \bibinfo{booktitle}{\emph{{IJCAI}}}.
\newblock


\bibitem[Yan et~al\mbox{.}(2019)]%
        {deepapf}
\bibfield{author}{\bibinfo{person}{Huan Yan}, \bibinfo{person}{Xiangning Chen},
  \bibinfo{person}{Chen Gao}, \bibinfo{person}{Yong Li}, {and}
  \bibinfo{person}{Depeng Jin}.} \bibinfo{year}{2019}\natexlab{}.
\newblock \showarticletitle{Deepapf: Deep attentive probabilistic factorization
  for multi-site video recommendation}.
\newblock  (\bibinfo{year}{2019}).
\newblock


\bibitem[Yao and Huang(2017)]%
        {FOCF}
\bibfield{author}{\bibinfo{person}{Sirui Yao} {and} \bibinfo{person}{Bert
  Huang}.} \bibinfo{year}{2017}\natexlab{}.
\newblock \showarticletitle{Beyond Parity: Fairness Objectives for
  Collaborative Filtering}. In \bibinfo{booktitle}{\emph{Advances in Neural
  Information Processing Systems}}, Vol.~\bibinfo{volume}{30}.
\newblock


\bibitem[Yu et~al\mbox{.}(2020)]%
        {yu2020tagnn}
\bibfield{author}{\bibinfo{person}{Feng Yu}, \bibinfo{person}{Yanqiao Zhu},
  \bibinfo{person}{Qiang Liu}, \bibinfo{person}{Shu Wu}, \bibinfo{person}{Liang
  Wang}, {and} \bibinfo{person}{Tieniu Tan}.} \bibinfo{year}{2020}\natexlab{}.
\newblock \showarticletitle{{TAGNN:} Target Attentive Graph Neural Networks for
  Session-based Recommendation}. In \bibinfo{booktitle}{\emph{{SIGIR}}}.
\newblock


\bibitem[Yu et~al\mbox{.}(2021a)]%
        {yu2021sept}
\bibfield{author}{\bibinfo{person}{Junliang Yu}, \bibinfo{person}{Hongzhi Yin},
  \bibinfo{person}{Min Gao}, \bibinfo{person}{Xin Xia},
  \bibinfo{person}{Xiangliang Zhang}, {and} \bibinfo{person}{Nguyen Quoc~Viet
  Hung}.} \bibinfo{year}{2021}\natexlab{a}.
\newblock \showarticletitle{Socially-Aware Self-Supervised Tri-Training for
  Recommendation}. In \bibinfo{booktitle}{\emph{{KDD}}}.
\newblock


\bibitem[Yu et~al\mbox{.}(2021b)]%
        {yu2021mhcn}
\bibfield{author}{\bibinfo{person}{Junliang Yu}, \bibinfo{person}{Hongzhi Yin},
  \bibinfo{person}{Jundong Li}, \bibinfo{person}{Qinyong Wang},
  \bibinfo{person}{Nguyen Quoc~Viet Hung}, {and} \bibinfo{person}{Xiangliang
  Zhang}.} \bibinfo{year}{2021}\natexlab{b}.
\newblock \showarticletitle{Self-Supervised Multi-Channel Hypergraph
  Convolutional Network for Social Recommendation}. In
  \bibinfo{booktitle}{\emph{{WWW}}}.
\newblock


\bibitem[Yu et~al\mbox{.}(2022)]%
        {yu2022simgcl}
\bibfield{author}{\bibinfo{person}{Junliang Yu}, \bibinfo{person}{Hongzhi Yin},
  \bibinfo{person}{Xin Xia}, \bibinfo{person}{Tong Chen},
  \bibinfo{person}{Lizhen Cui}, {and} \bibinfo{person}{Nguyen Quoc~Viet Hung}.}
  \bibinfo{year}{2022}\natexlab{}.
\newblock \showarticletitle{Are Graph Augmentations Necessary? Simple Graph
  Contrastive Learning for Recommendation}. In
  \bibinfo{booktitle}{\emph{{SIGIR}}}.
\newblock


\bibitem[Zehlike et~al\mbox{.}(2021)]%
        {zehlike2021fairness}
\bibfield{author}{\bibinfo{person}{Meike Zehlike}, \bibinfo{person}{Ke Yang},
  {and} \bibinfo{person}{Julia Stoyanovich}.} \bibinfo{year}{2021}\natexlab{}.
\newblock \showarticletitle{Fairness in ranking: A survey}.
\newblock \bibinfo{journal}{\emph{arXiv preprint arXiv:2103.14000}}
  (\bibinfo{year}{2021}).
\newblock


\bibitem[Zhang et~al\mbox{.}(2021b)]%
        {ZhangYZC021}
\bibfield{author}{\bibinfo{person}{Shengyu Zhang}, \bibinfo{person}{Dong Yao},
  \bibinfo{person}{Zhou Zhao}, \bibinfo{person}{Tat{-}Seng Chua}, {and}
  \bibinfo{person}{Fei Wu}.} \bibinfo{year}{2021}\natexlab{b}.
\newblock \showarticletitle{CauseRec: Counterfactual User Sequence Synthesis
  for Sequential Recommendation}. In \bibinfo{booktitle}{\emph{{SIGIR} 2021}}.
  \bibinfo{pages}{367--377}.
\newblock


\bibitem[Zhang et~al\mbox{.}(2019)]%
        {rs_survey}
\bibfield{author}{\bibinfo{person}{Shuai Zhang}, \bibinfo{person}{Lina Yao},
  \bibinfo{person}{Aixin Sun}, {and} \bibinfo{person}{Yi Tay}.}
  \bibinfo{year}{2019}\natexlab{}.
\newblock \showarticletitle{Deep Learning Based Recommender System: {A} Survey
  and New Perspectives}.
\newblock \bibinfo{journal}{\emph{{ACM} Comput. Surv.}} \bibinfo{volume}{52},
  \bibinfo{number}{1} (\bibinfo{year}{2019}), \bibinfo{pages}{5:1--5:38}.
\newblock


\bibitem[Zhang et~al\mbox{.}(2021a)]%
        {zhang2021causal}
\bibfield{author}{\bibinfo{person}{Yang Zhang}, \bibinfo{person}{Fuli Feng},
  \bibinfo{person}{Xiangnan He}, \bibinfo{person}{Tianxin Wei},
  \bibinfo{person}{Chonggang Song}, \bibinfo{person}{Guohui Ling}, {and}
  \bibinfo{person}{Yongdong Zhang}.} \bibinfo{year}{2021}\natexlab{a}.
\newblock \showarticletitle{Causal intervention for leveraging popularity bias
  in recommendation}. In \bibinfo{booktitle}{\emph{Proceedings of the 44th
  International ACM SIGIR Conference on Research and Development in Information
  Retrieval}}. \bibinfo{pages}{11--20}.
\newblock


\bibitem[Zhao et~al\mbox{.}(2021)]%
        {zhao2021recbole}
\bibfield{author}{\bibinfo{person}{Wayne~Xin Zhao}, \bibinfo{person}{Shanlei
  Mu}, \bibinfo{person}{Yupeng Hou}, \bibinfo{person}{Zihan Lin},
  \bibinfo{person}{Yushuo Chen}, \bibinfo{person}{Xingyu Pan},
  \bibinfo{person}{Kaiyuan Li}, \bibinfo{person}{Yujie Lu},
  \bibinfo{person}{Hui Wang}, \bibinfo{person}{Changxin Tian},
  \bibinfo{person}{Yingqian Min}, \bibinfo{person}{Zhichao Feng},
  \bibinfo{person}{Xinyan Fan}, \bibinfo{person}{Xu Chen},
  \bibinfo{person}{Pengfei Wang}, \bibinfo{person}{Wendi Ji},
  \bibinfo{person}{Yaliang Li}, \bibinfo{person}{Xiaoling Wang}, {and}
  \bibinfo{person}{Ji-Rong Wen}.} \bibinfo{year}{2021}\natexlab{}.
\newblock \showarticletitle{RecBole: Towards a Unified, Comprehensive and
  Efficient Framework for Recommendation Algorithms}. In
  \bibinfo{booktitle}{\emph{{CIKM}}}.
\newblock


\bibitem[Zheng et~al\mbox{.}(2021)]%
        {zheng2021disentangling}
\bibfield{author}{\bibinfo{person}{Yu Zheng}, \bibinfo{person}{Chen Gao},
  \bibinfo{person}{Xiang Li}, \bibinfo{person}{Xiangnan He},
  \bibinfo{person}{Yong Li}, {and} \bibinfo{person}{Depeng Jin}.}
  \bibinfo{year}{2021}\natexlab{}.
\newblock \showarticletitle{Disentangling user interest and conformity for
  recommendation with causal embedding}. In
  \bibinfo{booktitle}{\emph{Proceedings of the Web Conference 2021}}.
  \bibinfo{pages}{2980--2991}.
\newblock


\bibitem[Zhu et~al\mbox{.}(2018b)]%
        {zhu2018person}
\bibfield{author}{\bibinfo{person}{Chen Zhu}, \bibinfo{person}{Hengshu Zhu},
  \bibinfo{person}{Hui Xiong}, \bibinfo{person}{Chao Ma}, \bibinfo{person}{Fang
  Xie}, \bibinfo{person}{Pengliang Ding}, {and} \bibinfo{person}{Pan Li}.}
  \bibinfo{year}{2018}\natexlab{b}.
\newblock \showarticletitle{Person-job fit: Adapting the right talent for the
  right job with joint representation learning}.
\newblock \bibinfo{journal}{\emph{{TMIS}}} (\bibinfo{year}{2018}).
\newblock


\bibitem[Zhu et~al\mbox{.}(2019)]%
        {dtcdr}
\bibfield{author}{\bibinfo{person}{Feng Zhu}, \bibinfo{person}{Chaochao Chen},
  \bibinfo{person}{Yan Wang}, \bibinfo{person}{Guanfeng Liu}, {and}
  \bibinfo{person}{Xiaolin Zheng}.} \bibinfo{year}{2019}\natexlab{}.
\newblock \showarticletitle{Dtcdr: A framework for dual-target cross-domain
  recommendation}. In \bibinfo{booktitle}{\emph{Proceedings of the 28th ACM
  International Conference on Information and Knowledge Management}}.
  \bibinfo{pages}{1533--1542}.
\newblock


\bibitem[Zhu et~al\mbox{.}(2018a)]%
        {dcdcsr}
\bibfield{author}{\bibinfo{person}{Feng Zhu}, \bibinfo{person}{Yan Wang},
  \bibinfo{person}{Chaochao Chen}, \bibinfo{person}{Guanfeng Liu},
  \bibinfo{person}{Mehmet Orgun}, {and} \bibinfo{person}{Jia Wu}.}
  \bibinfo{year}{2018}\natexlab{a}.
\newblock \showarticletitle{A deep framework for cross-domain and cross-system
  recommendations}. In \bibinfo{booktitle}{\emph{IJCAI}}.
\newblock


\bibitem[Zhu et~al\mbox{.}(2022)]%
        {bars}
\bibfield{author}{\bibinfo{person}{Jieming Zhu}, \bibinfo{person}{Kelong Mao},
  \bibinfo{person}{Quanyu Dai}, \bibinfo{person}{Liangcai Su},
  \bibinfo{person}{Rong Ma}, \bibinfo{person}{Jinyang Liu},
  \bibinfo{person}{Guohao Cai}, \bibinfo{person}{Zhicheng Dou},
  \bibinfo{person}{Xi Xiao}, {and} \bibinfo{person}{Rui Zhang}.}
  \bibinfo{year}{2022}\natexlab{}.
\newblock \showarticletitle{BARS: Towards Open Benchmarking for Recommender
  Systems}.
\newblock \bibinfo{journal}{\emph{arXiv preprint arXiv:2205.09626}}
  (\bibinfo{year}{2022}).
\newblock


\bibitem[Zhu et~al\mbox{.}(2021)]%
        {zhu2021learning}
\bibfield{author}{\bibinfo{person}{Yongchun Zhu}, \bibinfo{person}{Ruobing
  Xie}, \bibinfo{person}{Fuzhen Zhuang}, \bibinfo{person}{Kaikai Ge},
  \bibinfo{person}{Ying Sun}, \bibinfo{person}{Xu Zhang}, \bibinfo{person}{Leyu
  Lin}, {and} \bibinfo{person}{Juan Cao}.} \bibinfo{year}{2021}\natexlab{}.
\newblock \showarticletitle{Learning to warm up cold item embeddings for
  cold-start recommendation with meta scaling and shifting networks}. In
  \bibinfo{booktitle}{\emph{Proceedings of the 44th International ACM SIGIR
  Conference on Research and Development in Information Retrieval}}.
  \bibinfo{pages}{1167--1176}.
\newblock


\end{thebibliography}

\end{document}